\documentclass[envcountsame,fleqn]{llncs}

\pdfoutput=1

\usepackage{mgilncs}
\usepackage{tikz}
\usepackage{ifthen}
\usepackage{permissive}
\usepackage{fixme}

\newif\iffull
\fulltrue
\title{Measuring Permissiveness in Parity Games:\\
    Mean-Payoff Parity Games Revisited\thanks{Sponsored by
      ANR-06-SETI-003 DOTS, and by 
    ESF-Eurocores LogICCC GASICS.}}

\author{Patricia Bouyer\addr{1}, Nicolas Markey\addr{1},
J\"org Olschewski\addr{2}, Michael Ummels\addr{1,3}}

\address{LSV, CNRS \& ENS Cachan, France \\
\email{\{bouyer,markey,ummels\}@lsv.ens-cachan.fr}}

\address{Lehrstuhl Informatik~7, RWTH Aachen University, Germany \\
\email{olschewski@automata.rwth-aachen.de}}

\address{LAMSADE, CNRS \& Universit\'e Paris-Dauphine, France}

\begin{document}

\maketitle

\setcounter{footnote}{0}

\begin{abstract}
  We study nondeterministic strategies in parity
  games with the aim of computing a \emph{most permissive} winning
  strategy. Following earlier work, we~measure permissiveness in terms
  of the \emph{average} number\slash weight of transitions blocked by
  a~strategy. 
  Using a translation into mean-payoff parity games,
  we~prove that deciding (the permissiveness~of)
  a~most permissive winning strategy is in $\NP\cap\coNP$.
  Along the way, we~provide a new study of mean-payoff~parity games.
  In~particular, we~give a new algorithm for solving these games,
  which beats all previously known algorithms for this problem.
\end{abstract}

\section{Introduction}

Games extend the usual semantics of finite automata from one to several players,
thus allowing to model interactions between agents acting on the progression of
the automaton. This has proved very useful in computer science, especially for
the formal verification of open systems interacting with their
environment~\cite{Tho02}. In~this setting, the aim is to synthesise a
controller under which the system behaves according to a
given specification, whatever the environment does.
Usually, this is modelled as a game between two players:
Player~1 represents the controller and Player~2 represents the environment.
The goal is then to find a~\emph{winning strategy} for Player~1, \ie
a~recipe stating how the system should react to any possible action of
the environment, in order to meet its specification.

In this paper, we consider \emph{multi-strategies} (or
\emph{non-deterministic strategies}, \cf~\cite{BJW02,BDMR09}) as a
generalisation of strategies: while strategies select only one possible action
to be played in response to the behaviour of the environment, multi-strategies
can retain several possible actions. Allowing several moves provides a~way to
cope with errors (\eg, actions being disabled for a short period, or timing
imprecisions in timed games).
Another quality of multi-strategies is their ability to be combined
with other multi-strategies, yielding a refined multi-strategy, which
is ideally winning for all of the original specifications.
This offers a modular approach for solving games. 

Classically, a strategy is more \emph{permissive} than another one if it
allows more behaviours. Under this notion, there does not need to exist
a most permissive winning strategy~\cite{BJW02}. Hence, we~follow
a~different approach, which is of a~quantitative nature: we provide
a~\emph{measure} that specifies \emph{how} permissive a given multi-strategy
is.
In~order to do so, we~consider \emph{weighted games}, where each edge is
equipped with a weight, which we treat as a \emph{penalty} that is
incurred when disallowing this edge. The penalty of a multi-strategy
is then defined to be the average sum of penalties incurred in each step
(in the limit). The lower this penalty is, the more permissive is the
given multi-strategy. Our~aim is to find one of the most permissive
multi-strategies achieving a given objective.

We deal with multi-strategies by transforming a game with penalties into a
\emph{mean-payoff game}~\cite{EM79,ZP96} with classical (deterministic)
strategies.
A~move in the latter game corresponds to a set of moves in the former, and is
assigned a (negative) \emph{reward} depending on the penalty of the
original move. The~penalty of a multi-strategy in the original game
equals the opposite of the payoff achieved by the corresponding strategy
in the mean-payoff game.
In~previous work, Bouyer et al.~\cite{BDMR09} introduced the notion of
penalties and showed how to compute permissive strategies \wrt reachability
objectives. We extend the study of~\cite{BDMR09} to parity
objectives. This is a significant extension because parity
objectives can express infinitary specifications.
Using the above transformation, we~reduce the problem of finding
a most permissive strategy in a parity game with penalties to that of
computing an
optimal strategy in a \emph{mean-payoff parity game}, which combines a
mean-payoff objective with a parity objective.

While mean-payoff parity games have already been
studied~\cite{CHJ05,BCHJ09,CD10}, we~propose a new proof that these
games are determined and that both players have optimal strategies.
Moreover, we prove that the second player does not only have an
optimal strategy with finite memory, but one that uses no memory at all.
Finally, we provide a new algorithm for computing the values of a mean-payoff
parity game, which is faster than the best known algorithms
for this problem; the running time is exponential in the number of priorities
and polynomial in the size of the game graph and the largest absolute weight.

In the second part of this paper, we present our results on parity games
with penalties. In particular, we prove the existence of most permissive
multi-strategies, and we show that the existence of a multi-strategy
whose penalty is less than a given threshold can be decided in
${\NP\cap\coNP}$. Finally, we adapt our deterministic algorithm
for mean-payoff parity games to parity games with penalties. Our
algorithm computes the penalties of a most permissive multi-strategy
in time exponential in the number of priorities and polynomial in the
size of the game graph and the largest penalty.

\subsubsection{Related work}
Penalties as we use them were defined in~\cite{BDMR09}. Other notions of
permissiveness have been defined in~\cite{BJW02,PR05}, but these notions
have the drawback that a most permissive strategy might not exist.
Multi-strategies have also been used for different purposes in~\cite{Lut08}.

The parity condition goes back to~\cite{EJ91,Mos91} and is fundamental
for verification.
Parity games admit optimal memoryless strategies for both players,
and the problem of deciding the winner is in ${\NP\cap\coNP}$.
As of this writing, it is not known whether parity games can be
solved in polynomial time; the best known algorithms run in time
polynomial in the size of the game graph but exponential in the number
of priorities.

Another fundamental class of games are games with quantitative objectives.
Mean-payoff games, where the aim is
to maximise the average weight of the transitions taken in a play,
are also in $\NP\cap\coNP$ and admit memoryless
optimal strategies \cite{EM79,ZP96}. The same is true for
\emph{energy games}, where the aim is to always keep the sum of the weights
above a given threshold \cite{CdAHS03,BFLMS08}.
In~fact, parity games can easily be reduced to mean-payoff 
or energy games~\cite{Jurdzinski98}.

Finally, several game models mixing several qualitative or quantitative
objectives have recently appeared in the literature: apart from mean-payoff
parity games, these include generalised parity games~\cite{CHP07},
energy parity games~\cite{CD10} and lexicographic mean-payoff (parity)
games~\cite{BCHJ09} as~well~as generalised energy and
mean-payoff games~\cite{CDHR10}.

\section{Preliminaries}
\label{sec-prelim}

A \emph{weighted game graph} is a tuple $G=(\V1,\V2,E,\weight)$,
where $\V{all}\coloneqq\V1\cupdot\V2$ is a finite set of
\emph{states}, 
$E\subseteq\V{all}\times\V{all}$ is the \emph{edge} or
\emph{transition relation}, and
$\weight\colon E\to\bbR$ is a function assigning a \emph{weight} to every
transition.
When weighted game graphs are subject to algorithmic processing, we
assume that these weights are integers; in~this case, we
set $W\coloneq\max\{1,\abs{\weight(e)}\mid e\in E\}$.
\iffull\par
Moreover, we define the \emph{size} of~$G$, denoted by~$\size{G}$, as
$\abs{\V{}}+\abs{E}\cdot\lceil\log_2 W\rceil$.
(Up to a linear factor, $\size{G}$~is the length of a
binary encoding of~$G$).
In the same spirit, the size~$\size{x}$
of a rational number~$x$ equals the total length of the binary
representations of its numerator and its denominator.
\fi

For $q\in\V{all}$, we write $qE$ for the set
$\{q'\in\V{all}\mid(q,q')\in E\}$ of
all successors of~$q$. We~require that $qE\neq\emptyset$ for all states
$q\in\V{all}$.
A subset $S\subseteq\V{all}$ is a \emph{subarena} of~$G$ if
$qE\cap S\neq\emptyset$ for all states $q\in S$.
If $S\subseteq\V{all}$ is a subarena of~$G$, then we can restrict~$G$ to
states in~$S$, in which case we obtain the weighted game graph
$G\restriction S\coloneq
(\V1\cap S,\V2\cap S,E\cap(S\times S),\weight\restriction S\times S)$.

A \emph{play} of~$G$ is an infinite sequence $\rho=\rho(0)\rho(1)\cdots\in
\V{}^\omega$ of states such that $(\rho(i),\rho(i+1))\in E$ for all $i\in\bbN$.
We denote by $\Out^{\G}(q)$ the set of all plays~$\rho$ with $\rho(0)=q$
and by~$\Inf(\rho)$ the set of states occurring infinitely often in~$\rho$.

A \emph{play prefix} or a \emph{history} $\gamma=\gamma(0)\gamma(1)\cdots
\gamma(n)\in \V{}^+$ is a finite, nonempty prefix of a play.
\iffull
For a play or a history~$\rho$ and $j<k\in\bbN$, we denote by
$\rho[j,k)\coloneqq\rho[j,k-1]\coloneqq\rho(j)\cdots\rho(k-1)$ its infix that
starts at position~$j$ and ends at position $k-1$; the play's suffix
$\rho(j)\rho(j+1)\cdots$ is denoted by $\rho[j,\infty)$.
\fi

\paragraph{Strategies}

A \emph{(deterministic) strategy} for \Pli in~$G$ is a function
$\sigma\colon \V{all}^* \V{i}\to\V{all}$ such that
$\sigma(\gamma q)\in qE$ for all $\gamma\in\V{}^*$ and $q\in\V{i}$.
A~strategy~$\sigma$ is \emph{memoryless} if
$\sigma(\gamma q)=\sigma(q)$ for all $\gamma\in\V{}^*$ and $q\in\V{i}$.
More generally, a strategy~$\sigma$ is \emph{finite-memory} if the equivalence
relation~${\sim}\subseteq\V{}^*\times\V{}^*$, defined by $\gamma_1\sim\gamma_2$
if and only if $\sigma(\gamma_1\cdot\gamma)=\sigma(\gamma_2\cdot\gamma)$ for
all $\gamma\in\V{}^*\V{i}$, has finite index.

We say that a play~$\rho$ of~$G$ is \emph{consistent} with a
strategy~$\sigma$
for \Pli if $\rho(k+1)=\sigma(\rho[0,k])$ for all $k\in\bbN$
with $\rho(k)\in\V{i}$, and  denote by
$\Out^{G}(\sigma,q_0)$ the set of all plays~$\rho$ of~$G$ that are
consistent with~$\sigma$ and start in $\rho(0)=q_0$.
Given a~strategy~$\sigma$ of \Pl1, a strategy~$\tau$
of \Pl2, and a state $q_0\in\V{all}$, there exists a unique play
$\rho\in\Out^G(\sigma,q_0)\cap\Out^G(\tau,q_0)$, which we denote by
$\rho^G(\sigma,\tau,q_0)$.

\paragraph{Traps and attractors}

Intuitively, a subarena $T\subseteq\V{all}$ of states is a \emph{trap} for
one of the two players if the other player can enforce that the play stays
in this set.
Formally, a~trap for \Pl2 (or simply a $2$-trap) is a subarena
$T\subseteq\V{all}$ such that
$qE\subseteq T$ for all states $q\in T\cap \V{2}$, and
$qE\cap T\neq\emptyset$ for all $q\in T\cap \V{1}$. A~trap for \Pl1
(or $1$-trap) is defined analogously.
\iffull
Note that if $T$~is n trap for \Pli in $G\restrict S$ and
$S$~is a trap for \Pl1 in~$G$, then $T$~is also a trap
for \Pli in~$G$.
\fi

If $T\subseteq\V{all}$ is not a trap for \Pl1, then \Pl1 has a strategy
to reach a position in $\V{all}\setminus T$. In~general, given a
subset $S\subseteq\V{all}$, we~denote by $\Attr_1^{G}(S)$ the set of states
from where \Pl1 can force a visit to~$S$.
\iffull
This set can be characterised as the limit of the 
sequence $(A_i)_{i\in\bbN}$ defined by $A^0 = S$ and
\[
A^{i+1} = A^i\cup \{q\in\V{1}\mid qE\cap A^i\neq\emptyset\} \cup
  \{q\in\V{2}\mid qE\subseteq A^i\}\,.
\]
From every state in~$\Attr_1^{G}(S)$, \Pl1 has a memoryless strategy~$\sigma$
that guarantees a visit to~$S$ in at most~$\abs{\V{all}}$ steps: the
strategy chooses for each state $q\in (A^i\setminus A^{i-1})\cap \V{1}$ a state
$p\in qE\cap A^{i-1}$ (which decreases the distance to~$S$ by~$1$).
We~call the set $\Attr_1^{G}(S)=\bigcup_{i\in\bbN}A_i$ the \emph{$1$-attractor
of~$S$} and $\sigma$ an \emph{attractor strategy for~$S$}.
\fi
The \emph{$2$-attractor} of a set~$S$, denoted by $\Attr_2^{G}(S)$,
and attractor strategies for \Pl2 are defined symmetrically. 
Notice that for any set~$S$, the~set $\V{}\setminus\Attr_1^G(S)$ is a $1$-trap,
and if $S$~is a subarena ($2$-trap), then $\Attr_1^G(S)$ is also a
subarena ($2$-trap).
Analogously, $\V{}\setminus\Attr_2^G(S)$ is a $2$-trap, and if $S$~is a
subarena ($1$-trap), then $\Attr_2^G(S)$ is also a subarena ($1$-trap).

\paragraph{Convention}
We often drop the superscript~$G$ from the expressions defined above,
if no confusion arises, \eg by writing $\Out(\sigma,q_0)$ instead of
$\Out^{G}(\sigma,q_0)$.

\section{Mean-payoff parity games}
\label{sec-mpp}

\iffull
In this first part of the paper, we show that mean-payoff parity games
are determined, that both players have
optimal strategies, that for \Pl2 even memoryless strategies suffice,
and that the value problem for mean-payoff parity games is in $\NP\cap\coNP$.
\fi
Furthermore, we present a deterministic algorithm which computes
the values in time exponential in the number of priorities, and runs 
in pseudo-polynomial time 
when the number of priorities is bounded.

\iffull
\subsection{Definitions}
\fi

Formally, a \emph{mean-payoff parity game} is a tuple $\G=(G,\col)$, where
$G$~is a~weighted game graph, and $\col\colon \V{all}\to\bbN$ is a priority
function assigning a \emph{priority} to every state.
A play $\rho=\rho(0)\rho(1)\cdots$ is \emph{parity-winning} if the minimal
priority occurring infinitely often in $\rho$ is even, \ie, if
$\min\{\col(q)\mid q\in\Inf(\rho)\}\equiv 0\pmod 2$. All notions that we
have defined for weighted game graphs carry over to mean-payoff parity games.
In~particular, a~play of~$\G$ is just a play of~$G$ and a strategy for \Pli
in~$\G$ is nothing but a strategy for \Pli in~$G$. Hence, we write
$\Out^{\G}(\sigma,q)$ for $\Out^{G}(\sigma,q)$, and so on. As for
weighted games graphs, we often omit the superscript if
$\G$~is clear from the context.
Finally, for a mean-payoff parity game~$\G=(G,\col)$ and a subarena~$S$
of~$G$, we~write $\G\restrict S$ for the mean-payoff parity game
$(G\restrict S, \col\restrict S)$.

We say that a mean-payoff parity game $\G=(G,\col)$ is a \emph{mean-payoff
game} if $\col(q)$ is even for all $q\in\V{all}$. In particular, given
a weighted game graph~$G$, we~obtain a mean-payoff game by assigning
priority~$0$ to all states. We denote this game by~$(G,0)$.

\iffull
If $\col(\V{all})\subseteq\{0,1\}$, then we say that $\G$~is a
\emph{mean-payoff B\"uchi game}; if $\col(\V{all})\subseteq\{1,2\}$, we
call it a \emph{mean-payoff co-B\"uchi game}. Hence, in a B\"uchi game \Pl1
needs to visit the set $\col^{-1}(0)$ infinitely often, whereas in a
co-B\"uchi game he has to visit the set $\col^{-1}(1)$ only
finitely often.
\fi

\iffull
For a play $\rho$ of~$\G$, we define its \emph{payoff} as
\[
\payoff^{\G}(\rho)=\begin{cases}
\displaystyle\liminf_{n\to\infty}\payoff^{\G}_n(\rho)
 & \text{if $\rho$ is parity-winning,} \\
-\infty	& \text{otherwise,}
\end{cases}
\]
where for $n\in\bbN$
\[
\payoff^{\G}_n(\rho)=\begin{cases}
\displaystyle\frac{1}{n}\sum^{n-1}_{i=0}\weight(\rho(i),\rho(i+1))
 & \text{if $n>0$,} \\
-\infty & \text{if $n=0$.}
\end{cases}
\]
\fi
If $\sigma$~is a strategy for \Pl1 in~$\G$, we define its \emph{value}
from $q_0\in\V{all}$ as
\iffull
\[\val^{\G}(\sigma,q_0)
=\inf\nolimits_{\tau}\payoff^{\G}(\rho(\sigma,\tau,q_0))
=\inf\{\payoff^{\G}(\rho)\mid\rho\in\Out^{\G}(\sigma,q_0)\},\]
where $\tau$~ranges over all strategies of \Pl2 in~$\G$.
\fi
Analogously,
\iffull
the value of a strategy~$\tau$ for \Pl2 from~$q_0$ is defined as
\[\val^{\G}(\tau,q_0)
=\sup\nolimits_{\sigma}\payoff^{\G}(\rho(\sigma,\tau,q_0))
=\sup\{\payoff^\G(\rho)\mid\rho\in\Out^\G(\tau,q_0)\},\]
where $\sigma$~ranges over all strategies of \Pl1 in~$\G$.
\fi
The \emph{lower} and \emph{upper value} of a state~$q_0\in\V{all}$
are defined by
\iffull
\begin{align*}
\lowval^{\G}(q_0)=\sup\nolimits_{\sigma}\val^{\G}(\sigma,q_0)
  && \text{and} &&
\upval^{\G}(q_0)=\inf\nolimits_{\tau}\val^{\G}(\tau,q_0),
\end{align*}
\fi
respectively.
Intuitively, $\lowval^{\G}(q_0)$ and $\upval^{\G}(q_0)$ are the maximal
(respectively minimal) payoff that \Pl1 (respectively \Pl2) can ensure
(in the limit). We~say that a~strategy~$\sigma$ of \Pl1 is \emph{optimal
from~$q_0$} if $\val^{\G}(\sigma,q_0)=\lowval^{\G}(q_0)$. Analogously,
we~call a strategy~$\tau$ of \Pl2 optimal from~$q_0$ if
$\val^{\G}(\tau,q_0)=\upval^{\G}(q_0)$. A~strategy is
\emph{(globally) optimal} if it is optimal from every state
$q\in\V{all}$.
It is easy to see that $\lowval^{\G}(q_0)\leq\upval^{\G}(q_0)$. If
$\lowval^{\G}(q_0)=\upval^{\G}(q_0)$, we say that $q_0$~has a \emph{value},
which we denote by $\val^\G(q_0)$.

\iffull
In the next section, we will see that mean-payoff games are
\emph{determined}, \ie, that every state has a value.
The \emph{value problem} is the following decision problem:
Given a mean-payoff parity game~$\G$ (with integral weights), a
designated state $q_0\in\V{all}$, and a number $x\in\bbQ$, decide
whether $\val^{\G}(q_0)\geq x$.
\fi

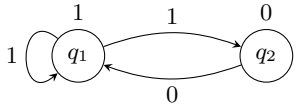
\begin{figure}
\centering
\begin{tikzpicture}[x=2.5cm,->,bend angle=20]
\node[pl1] (0) at (0,0) [label={above:$1$}] {$q_1$};
\node[pl1] (1) at (1,0) [label={above:$0$}] {$q_2$};

\draw (0) to [loop,out=140,in=-140,looseness=4] node[left] {$1$} (0);
\draw (0) to [bend left] node[above] {$1$} (1);
\draw (1) to [bend left] node[below] {$0$} (0);
\end{tikzpicture}
\caption{\label{fig:inf-mem}A mean-payoff parity game for which infinite
memory is necessary}
\end{figure}

\begin{example}\label{ex:inf-mem}
Consider the mean-payoff parity game~$\G$ depicted in \cref{fig:inf-mem},
where a state or an edge is labelled with its priority, respectively
weight; all states belong to \Pl1.
Note that $\val^{\G}(q_1)=1$ since \Pl1 can delay visiting~$q_2$
longer and longer while still ensuring that this vertex is seen
infinitely often. However, there is no finite-memory strategy
that achieves this value.
\iffull\par
Let $\sigma$ be a finite-memory
strategy of \Pl1 in~$\G$, and let $\rho$ be the
unique play of~$\G$ that starts in~$q_1$ and is consistent
with~$\sigma$. Assume furthermore that $\rho$~visits~$q_2$
infinitely often (otherwise $\val^{\G}(\sigma,q_1)=-\infty$). Then
$\rho={q_1}^{k_1}q_2 {q_1}^{k_2} q_2\cdots$, where each
$k_i\in\bbN\setminus\{0\}$. Since $\sigma$~is a finite-memory strategy,
there exists $m\in\bbN$ such that $k_i\leq m$ for all $i\in\bbN$.
Hence, $\val^{\G}(\sigma,q_q)=\payoff(\rho)\leq m/(m+1)<1$.
\fi
\end{example}

\iffull
\subsection{Strategy complexity}
\fi

\iffull
It follows from Martin's determinacy theorem \cite{Martin75} that
mean-payoff parity games are determined.
\fi
Moreover, Chatterjee et
al.~\cite{CHJ05} gave an algorithmic proof for the existence of optimal
strategies. Finally, it can be shown that for every
$x\in\bbR\cup\{-\infty\}$ the set
${\{\rho\in\V{}^\omega\mid\payoff(\rho)\geq x\}}$ is closed under
\emph{combinations}. By Theorem~4 in \cite{Kopczynski06}, this property
implies that \Pl2 even has a memoryless optimal strategy.
\iffull
We give here a purely inductive proof of these facts that does not
rely on Martin's theorem.
We start by proving that \Pl1 has an optimal strategy in games where
\Pl2 is absent.

\begin{lemma}
\label{lem:mpp-pl1}
Let $\G$ be a mean-payoff parity game with $\V2=\emptyset$.
Then \Pl1 has an optimal strategy in~$\G$.
\end{lemma}

\begin{proof}
It suffices to construct for each $q_0\in\V{}$ a strategy~$\sigma$ with
$\val^{\G}(\sigma,q_0)\geq\val^{\G}(q_0)$.
If $\val^{\G}(q_0)=-\infty$, we can choose an arbitrary
strategy~$\sigma$. Otherwise, by the definition of $\val^{\G}(q_0)$, for
each $\epsilon>0$ there exists a play $\rho_\epsilon\in\Out^\G(q_0)$ with
$\payoff(\rho_\epsilon)\geq\val^{\G}(q_0)-\epsilon$. Consider the sets
$\Inf(\rho_\epsilon)$ of states occurring infinitely often
in~$\rho_\epsilon$. Since there are only finitely many such sets, we
can find a set $P\subseteq\V{all}$ such that for each $\epsilon>0$
there exists $0<\epsilon'<\epsilon$ with $P=\Inf(\rho_{\epsilon'})$.
Let $q_{\min}\in P$ be a vertex of lowest priority. (This priority must be
even since each~$\rho_\epsilon$ fulfils the parity condition).

Let $\sigma_1$ be an optimal memoryless strategy in the mean-payoff
game $\G_P=(G\restrict P,0)$ (the strategy~$\sigma_1$ just leads the
play to a simple cycle with maximum average weight),
and let $\sigma_2$ be the memoryless attractor strategy in the game~$\G_P$
that ensures a visit to~$q_{\min}$ from all states $q\in P$; we
extend both strategies to a strategy in~$\G$ by combining them with a
memoryless attractor strategy for~$P$. (In particular, $\sigma_2$~enforces
a visit to~$q_{\min}$ from~$q_0$.)
Note that $\val^{\G_P}(q)\geq\val^{\G}(q_0)$ for all $q\in P$
since each of the plays~$\rho_{\epsilon'}$ visits each vertex in~$P$
and has payoff $\geq\val^{\G}(q_0)-\epsilon'$.

\Pl1's optimal strategy~$\sigma$ is played in rounds: in the $i$th
round, \Pl1 first forces a visit to~$q_{\min}$ by playing according
to~$\sigma_2$; once $q_{\min}$~has been visited, \Pl1 plays
$\sigma_1$ for $i$~steps before proceeding to the next round.
Note that $\val^{\G_P}(\sigma,q_{\min})=
\val^{\G_P}(\sigma_1,q_{\min})$. Moreover, the unique play
$\rho\in\Out^{\G}(\sigma,q_0)$ satisfies $q_{\min}\in\Inf(\rho)\subseteq P$
and therefore fulfils the parity condition. To sum up, we have
$\val^{\G}(\sigma,q_0)=
\val^{\G}(\sigma,q_{\min})=
\val^{\G_P}(\sigma,q_{\min})=
\val^{\G_P}(\sigma_1,q_{\min})=
\val^{\G_P}(q_{\min})\geq\val^{\G}(q_0)$.\qed
\end{proof}

Using \cref{lem:mpp-pl1}, we can prove that mean-payoff-parity games are not
only determined, but also that \Pl1 has an optimal strategy and that \Pl2 has
a memoryless optimal strategy.

We use the \emph{loop factorisation} technique (\cf~\cite{Zielonka04}):
Let $\gamma$ be a play prefix and let $\hat{q}\in\V{all}$.
The \emph{loop factorisation of~$\gamma$ relative to~$\hat{q}$} is the unique
factorisation of the form $\gamma=\gamma_0\gamma_1\cdots\gamma_l$, where
$\gamma_0$ does not contain $\hat{q}$, and each factor $\gamma_i$,
$1\leq i\leq l$,
is of the form $\gamma_i=\hat{q}\cdot\gamma_i'$ where $\gamma_i'$ does not
contain~$\hat{q}$.
Analogously, for a play $\rho$ which has infinitely many occurrences
of~$\hat{q}$ the \emph{loop factorisation of~$\rho$ relative to~$\hat{q}$} is
the unique factorisation $\rho=\gamma_0\gamma_1\cdots$ where each $\gamma_i$
has the same properties as in the above case.

For a state~$\hat{q}$ with $m$~successors,
$\hat{q}E=\{q_1,\ldots,q_m\}$,
we define an operator $\pi_i\colon\V{all}^*\to\V{all}^*$ for each
$1\leq i\leq m$ by setting
\[
\pi_i(\gamma)\coloneqq\begin{cases}
	\gamma	& \text{if either $\gamma=\hat{q}q_i\gamma'$ for some $\gamma'\in\V{all}^*$ or $\gamma=q_i=\hat{q}$,} \\
	\epsilon	& \text{otherwise.}
\end{cases}
\]
The operator~$\pi_i$ induces another operator $\Pi_i\colon
\V{all}^*\to\V{all}^*$ by setting
\[\Pi_i(\gamma)=\Pi_i(\gamma_0)\Pi_i(\gamma_1)\cdots\Pi_i(\gamma_l),\]
where $\gamma=\gamma_0\gamma_1\cdots\gamma_l$ is the loop factorisation
of~$\gamma$ relative to~$\hat{q}$. The operator~$\Pi_i$~operates on play
prefixes, but it can easily be extended to operate on infinite plays with
infinitely many occurrences of~$\hat{q}$.

\fi

\begin{theorem}
\label{thm:mpp-main}
Let $\G$ be a mean-payoff parity game.
\begin{enumerate}
	\item $\G$~is determined;
	\item \Pl1 has an optimal strategy in $\G$;
	\item \Pl2 has a memoryless optimal strategy in $\G$.
\end{enumerate}
\end{theorem}

\iffull

\begin{proof}
We proceed by an induction over the size of
$S\coloneq\{{q\in\V2}\mid {\abs{qE}>1}\}$, the set of all
\Pl2 states with more than one successor.
If $S=\emptyset$, all statements follow from \cref{lem:mpp-pl1}.
Let 1.--3.\ be fulfilled for all games with $\abs{S}<n$ and let $\G=(G,\col)$
be a mean-payoff parity game with $\abs{S}=n$. We prove that the statements
also hold for~$\G$. Let $\hat{q}\in S$ with $\hat{q}E=\{q_1,\ldots,q_m\}$.
For each $1\leq j\leq m$, we define a new game $\G_j=(G_j,\col)$ by setting
$E_j=E\setminus(\{\hat{q}\}\times \V{all})\cup\{(\hat{q},q_j)\}$, and
$G_j=(\V1,\V2,E_j,\weight\restriction E_j)$.
Note that the induction hypothesis applies to each~$\G_j$.
\Wlg assume that $\val^{\G_1}(\hat{q})\leq\val^{\G_j}(\hat{q})$ for all
$1\leq j\leq m$.
We will construct a memoryless strategy~$\tau$ for \Pl2 and a strategy~$\sigma$
for \Pl1 such that $\val^{\G}(\tau,q_0)\leq\val^{\G_1}(q_0)$ and
$\val^{\G}(\sigma,q_0)\geq\val^{\G_1}(q_0)$ for every $q_0\in\V{all}$.
Hence,
\begin{align*}
\val^{\G_1}(q_0) \leq \val^{\G}(\sigma,q_0) \leq \lowval^{\G}(q_0) \leq \upval^{\G}(q_0) \leq \val^{\G}(\tau,q_0) \leq \val^{\G_1}(q_0),
\end{align*}
and all these numbers are equal. In particular, we have
$\val^{\G}(q_0)=\lowval^{\G}(q_0)=\upval^{\G}(q_0)$,
$\val^{\G}(\sigma,q_0)=\val^{\G}(q_0)$ and $\val^{\G}(\tau,q_0)
=\val^{\G}(q_0)$,
which proves 1.--3.

By the induction hypothesis, \Pl2 has a memoryless optimal strategy~$\tau$
in~$\G_1$.
Clearly, $\tau$~is also a memoryless strategy for \Pl2 in~$\G$, and
$\val^{\G}(\tau,q_0)=\val^{\G_1}(\tau,q_0)=\val^{\G_1}(q_0)$ for all
$q_0\in\V{all}$.

It remains to construct a strategy~$\sigma$ for \Pl1 in~$\G$ such that
$\val^{\G}(\sigma,q_0)\geq\val^{\G_1}(q_0)$ for all $q_0\in\V{all}$.

First, we devise a strategy~$\hat{\sigma}$ such that
$\val^{\G}(\hat{\sigma},\hat{q})\geq\val^{\G_1}(\hat{q})$.
If $\val^{\G_1}(\hat{q})=-\infty$, we can take an arbitrary strategy.
Hence, assume that $\val^{\G_1}(\hat{q})$ is finite.
By the induction hypothesis, for each $j=1,\ldots,m$ there exists a
strategy~$\sigma_j$ for \Pl1 in~$\G_j$ with
$\val^{\G_j}(\sigma_j,\hat{q})=\val^{\G_j}(\hat{q})$.
We define $\hat{\sigma}$ to be the \emph{interleaving strategy}, defined by
\begin{align*}
\hat{\sigma}(\gamma)=\hat{\sigma}(\gamma_0\cdots\gamma_l)=
\begin{cases}
\sigma_1(\Pi_1(\gamma))	& \text{if $\gamma_l=\hat{q}q_1\gamma'$
  for some $\gamma'\in\V{all}^*$,} \\
\qquad\vdots & \qquad\vdots \\
\sigma_m(\Pi_m(\gamma)) & \text{if $\gamma_l=\hat{q}q_m\gamma'$
  for some $\gamma'\in\V{all}^*$,}
\end{cases}
\end{align*}
for all play prefixes~$\gamma$ whose loop factorisation relative to~$\hat{q}$
equals $\gamma_0\cdots\gamma_l$.
We~claim that $\val^{\G}(\hat{\sigma},\hat{q})\geq\val^{\G_1}(\hat{q})$.

Let $\rho\in\Out^\G(\hat{\sigma},\hat{q})$. If $\rho$~has only finitely many
occurrences of $\hat{q}$, then $\rho$~is equivalent to a play in~$\G_j$ that is
consistent with~$\sigma_j$ for some~$j$. Since
$\val^{G_j}(\hat{q})\geq\val^{G_1}(\hat{q})$ and $\sigma_j$ is
optimal, $\payoff(\rho)\geq\val^{G_1}(\hat{q})$, and we are done.
Otherwise, consider the loop factorisation $\rho=\gamma_0\gamma_1\cdots$ and
set
\[
\Gamma=\{j\in \{1,\ldots,m\}\mid \text{$\gamma_i\cdot\hat{q}$ is a loop in
$\G_j$ for infinitely many~$i\in\bbN$}\}.
\]
Since the mean-payoff parity condition is prefix-independent, we can
assume \wlg that every loop in~$\rho$ is
a loop in~$\G_j$ for $j\in\Gamma$. For each $j\in\Gamma$, denote by
$\rho_j=\Pi_j(\rho)$ the corresponding play in~$\G_j$. By definition
of~$\hat{\sigma}$, we have $\rho_j\in\Out^{\G_j}(\sigma_j,\hat{q})$ for
each $j\in\Gamma$. Since $\val^{\G_1}(\hat{q})$ is finite
and $\val^{\G_1}(\hat{q})\leq\val^{\G_j}(\hat{q})$, each~$\rho_j$ fulfils
the parity condition. As the minimal priority occurring infinitely often
in~$\rho$ also occurs infinitely often in one~$\rho_j$, this implies that
$\rho$~fulfils the parity condition.

We claim that for each $n>0$, $\payoff_n(\rho)$ is a weighted average of
$\payoff_{n_j}(\rho_j)$ for some $n_j>0$. To see this, consider the
loop factorisation $\gamma'_0\cdots\gamma'_k$ of $\rho[0,n]$. (Note
that $\gamma'_i=\gamma_i$ for all $i<k$.) For each $j\in\Gamma$, set
\[
n_j=\begin{cases}
\abs{\Pi_j(\rho[0,n])} -1 &
\text{if $\gamma'_k$ is a history of~$\G_j$ and either
$\gamma'_k\neq\hat{q}$ or $q_j=\hat{q}$.} \\
\abs{\Pi_j(\rho[0,n])} & \text{otherwise.}
\end{cases}
\]
Intuitively, $n_j$~is the number of transitions in~$\rho[0,n]$ that
correspond to a transition in~$\rho_j$. Hence,
\[\{(\rho(i),\rho(i+1))\mid 0\leq i<n\}=\bigcup_{j\in\Gamma}
\{(\rho_j(i),\rho_j(i+1))\mid 0\leq i<n_j\}.\]
In particular, $\sum_{j\in\Gamma} n_j=n$ and $\sum_{j\in\Gamma}
n_j/n=1$. We have
\begin{align*}
\payoff_n(\rho)
&= \frac{1}{n}\,\sum_{i=0}^{n-1}\weight(\rho(i),\rho(i+1)) \\
	\noalign{\pagebreak[1]}
&= \frac{1}{n}\sum_{\substack{j\in\Gamma \\ n_j>0}}\sum_{i=0}^{n_j-1}
\weight(\rho_j(i),\rho_j(i+1)) \\
	\noalign{\pagebreak[1]}
&= \sum_{\substack{j\in\Gamma \\ n_j>0}}\frac{n_j}{n}\cdot\frac{1}{n_j}
\sum_{i=0}^{n_j-1}\weight(\rho_j(i),\rho_j(i+1)) \\
	\noalign{\pagebreak[1]}
&= \sum_{\substack{j\in\Gamma \\ n_j>0}}
\frac{n_j}{n}\cdot\payoff_{n_j}(\rho_j).
\end{align*}
Since a weighted average is always bounded from below by the minimum element,
we can conclude that
\[\payoff_n(\rho)\geq
\min_{\substack{j\in\Gamma \\ n_j>0}}\payoff_{n_j}(\rho_j)\geq
\min_{j\in\Gamma}\payoff_{n_j}(\rho_j).\]
Taking the lower limit on both sides, we obtain
\begin{align*}
\payoff(\rho)
&= \liminf_{n\to\infty}\payoff_n(\rho) \\
	\noalign{\pagebreak[1]}
&\geq\liminf_{n\to\infty}\min_{j\in\Gamma}\payoff_{n_j}(\rho_j) \\
	\noalign{\pagebreak[1]}
&=\min_{j\in\Gamma}\liminf_{n\to\infty}\payoff_{n_j}(\rho_j) \\
	\noalign{\pagebreak[1]}
&=\min_{j\in\Gamma}\liminf_{n_j\to\infty}\payoff_{n_j}(\rho_j) \\
	\noalign{\pagebreak[1]}
&=\min_{j\in\Gamma}\payoff(\rho_j).
\end{align*}
Since each~$\rho_j$ is consistent with~$\sigma_j$ and $\sigma_j$~is
optimal, we have $\payoff(\rho_j)\geq\val^{\G_j}(\hat{q})
\geq\val^{\G_1}(\hat{q})$ for each $j\in\Gamma$ and therefore also $\payoff(\rho)\geq\val^{\G_1}(\hat{q})$. Since this holds for
all $\rho\in\Out^{\G}(\hat{\sigma},\hat{q})$, we can conclude that
$\val^{\G}(\hat{\sigma},\hat{q})\geq\val^{\G_1}(\hat{q})$.

Finally, we construct a strategy~$\sigma$ for \Pl1 in~$\G$ such that
$\val^{\G}(\sigma,q_0)\geq\val^{\G_1}(q_0)$ for all $q_0\in\V{all}$.
Let
\begin{align*}
\sigma(\gamma)=\begin{cases}
 \sigma_1(\gamma) & \text{if $\hat{q}$ does not occur in $\gamma$,} \\
 \hat{\sigma}(\hat{q}\gamma_2) & \text{if $\gamma=\gamma_1\hat{q}\gamma_2$ 
  with $\gamma_1\in(\V{all}\setminus\{\hat{q}\})^*$.}
\end{cases}
\end{align*}
Then for each play $\rho\in\Out^{\G}(\sigma,q_0)$ where $\hat{q}$ does not
occur, it holds $\payoff^\G(\rho)=\payoff^{\G_1}(\rho)\geq\val^{\G_1}(\sigma_1,q_0)=
\val^{\G_1}(q_0)$.
If $\hat{q}$ occurs in at least one play consistent with~$\sigma$, then in
the game~$\G_1$ (where $\sigma_1$ is optimal), we have
$\val^{\G_1}(q_0)=\val^{\G_1}(\sigma_1,q_0)\leq\val^{\G_1}(\hat{q})$.
Hence, for each play $\rho\in\Out^{\G}(\sigma,q_0)$ where $\hat{q}$ occurs
(say at position~$j$), it holds
$\payoff^\G(\rho)=\payoff^{\G}(\rho[j,\infty))\geq
\val^{\G}(\hat{\sigma},\hat{q})\geq\val^{\G_1}(\hat{q})\geq\val^{\G_1}(q_0)$.
Altogether we have $\payoff^\G(\rho)\geq\val^{\G_1}(q_0)$ for every play
$\rho\in\Out^{\G}(\sigma,q_0)$ and therefore
$\val^{\G}(\sigma,q_0)\geq\val^{\G_1}(q_0)$.\qed
\end{proof}

\fi

A consequence of the proof of
\iffull
\cref{lem:mpp-pl1,thm:mpp-main}
\fi
is that
each value of a~mean-payoff parity game is either $-\infty$ or equals one of
the values of a~mean-payoff game played on the same weighted graph (or
a~subarena of it). Since optimal memoryless strategies exist in
mean-payoff games~\cite{EM79}, the values of a~mean-payoff game with integral
weights are rational numbers of the form~$r/s$ with
$\abs{r}\leq\abs{\V{}}\cdot W$ and $\abs{s}\leq\abs{\V{}}$.
Consequently, this property holds for the (finite) values of a mean-payoff
parity game as well.

\iffull

While \cref{ex:inf-mem} demonstrates that an optimal strategy of \Pl1
requires infinite memory in general, this is not the case for
mean-payoff co-B\"uchi games, where both players have memoryless
optimal strategies. This can be seen by applying Theorem~2 of
\cite{GZ04} or by an inductive proof, which we provide here.

\begin{theorem}\label{thm:mpp-buechi-optimal}
Let $\G$ be a mean-payoff co-B\"uchi game. Then \Pl1 has a memoryless
optimal strategy from every state $q_0\in\V{all}$.
\end{theorem}

\begin{proof}
The proof is by induction over the number $\abs{\V{}}=n$ of states in~$\G$.
For $n=1$, the statement is trivially fulfilled.
Now let $n>1$, $q_0\in\V{}$, and assume that the statement is true for all
games with less than $n$~states. Define
$\V{}'=\V{}\setminus\Attr_2(\col^{-1}(1))$. If $\V{}'=\emptyset$, then \Pl2
can force visiting~$\col^{-1}(1)$ infinitely often by playing a
memoryless attractor strategy. Hence, $\val^{\G}(q_0)=-\infty$, and
every memoryless strategy of \Pl1 is optimal.
In the following, assume that $\V{}'\neq\emptyset$.
Consider the game $\G'\coloneqq\G\restrict \V{}'$,
which is a mean-payoff game, and set
\[S\coloneqq\{q\in\V{}'\mid\val^{\G'}(q)\geq\val^{\G}(q_0)\}.\]
Note that $S$~is a trap for \Pl2 both in~$\G'$ and in~$\G$
(since $\V{}'$~is a 2-trap in~$\G$).
We claim that $S\neq\emptyset$.
Towards a contradiction, assume that $S=\emptyset$, \ie,
$\val^{\G'}(q)<\val^{\G}(q_0)$ for all $q\in\V{}'$, and let $\tau$
be an optimal memoryless strategy for \Pl2 in~$\G'$. We extend $\tau$ to
a strategy in~$\G$ by combining it with a memoryless attractor strategy
for $\col^{-1}(1)$ on $\Attr_2(\col^{-1}(1))$. Let
$\rho\in\Out^{\G}(\tau,q_0)$ and $m\coloneq\max_{q\in\V{}'}\val^{\G'}(q)$.
Either $\rho$ visits $\Attr_2(\col^{-1}(1))$ and therefore also
$\col^{-1}(1)$ infinitely often, in which case
$\payoff(\rho)=-\infty<m$, or
$\rho[i,\infty)$ is a play of $\G'$ for some $i\in\bbN$, in which case
$\payoff(\rho)=\payoff(\rho[i,\infty))\leq\val^{\G'}(\rho(i))\leq m$.
Hence, $\val^{\G}(q_0)\leq\val^{\G}(\tau,q_0)\leq m<\val^{\G}(q_0)$, a
contradiction.

Now, let $\sigma'$ be a memoryless optimal strategy of \Pl1 in~$\G'$.
By the definition of~$S$, we have $\val^{\G'}(\sigma',q)\geq\val^\G(q_0)$
for all ${q\in S}$. Moreover, $\sigma'$~induces a memoryless
strategy~$\sigma_S$ in $\G\restrict S$ such that
$\val^{\G\restrict S}(\sigma_S,q)=\val^{\G'}(\sigma',q)\geq
\val^\G(q_0)$ for all $q\in S$. Let $A=\Attr_1^\G(S)$.
We extend~$\sigma_S$ to a memoryless strategy~$\sigma_A$
in~$\G\restrict A$ by combining it with a memoryless attractor
strategy for~$S$ on~$A\setminus S$. It follows that
$\val^{\G\restrict A}(\sigma_A,q)\geq
\val^\G(q_0)$ for all $q\in\Attr_1(S)$.
If $q_0\in\Attr_1(S)$, we are done. Otherwise, $q_0\in T\coloneq
\V{all}\setminus A$. Since $S\neq\emptyset$, the game
$\G\restrict T$ has less states than~$\G$, and by the induction hypothesis,
\Pl1 has a memoryless optimal strategy~$\sigma_T$ from~$q_0$
in~$\G\restrict T$. Note that, since $T$~is a trap for \Pl1, we~have
$\val^{\G\restrict T}(\sigma_T,q_0)=\val^{\G\restrict T}(q_0)
\geq\val^{\G}(q_0)$.
Let $\sigma$ be the union of $\sigma_A$ and~$\sigma_T$,
which is a memoryless strategy in~$\G$. We claim that $\sigma$~is optimal
from~$q_0$ in~$\G$. Let $\rho\in\Out^{\G}(\sigma,q_0)$. If $\rho$~stays
in~$T$, it is consistent with~$\sigma_T$ and must have
payoff at least $\val^{\G\restrict T}(\sigma_T,q_0)\geq\val^{\G}(q_0)$.
Otherwise, there exists $i\in\bbN$ such that $\rho(i)\in A$ and
$\rho[i,\infty)$ is consistent with~$\sigma_A$, which implies
$\payoff(\rho)=\payoff(\rho[i,\infty))\geq
\val^{\G\restrict A}(\sigma_A,\rho(i))\geq\val^\G(q_0)$.\qed
\end{proof}

\fi

\iffull
\subsection{Computational complexity}

In this section, we prove that the value problem for mean-payoff parity
games lies in $\NP\cap\coNP$. Although this has already been proved by
Chatterjee and Doyen~\cite{CD10}, our proof has the advantage that it
works immediately on mean-payoff parity games,
and not on energy parity games as in \cite{CD10}.
\fi

\iffull
In order to put the value problem for mean-payoff parity games into \coNP,
we first show that the value can be decided in polynomial time in games
where \Pl2 is absent.


\begin{proposition}\label{prop:mpp-pl1-ptime}
The problem of deciding, given	a mean-payoff parity game~$\G$ with $\V{2}=\emptyset$, a state $q_0\in\V{all}$, and $x\in\bbQ$,
whether $\val^{\G}(q_0)\geq x$, is in~\PTime.
\end{proposition}

\begin{proof}
Deciding whether $\val^{\G}(q_0)\geq x$ is achieved by
\cref{alg:pl1-value},
\begin{algorithm}
\vspace*{.8ex}
\begin{tabbing}
\hspace*{1em}\=\hspace{1em}\=\hspace{1em}\=\hspace{1em}\=
\hspace{1em}\=\hspace{1em}\= \kill

\emph{Input:} mean-payoff parity game $\G$ with~$\V2=\emptyset$,
$q_0\in\V{}$, $x\in\bbQ$. \\ 
\emph{Output:} whether $\val^{\G}(q_0)\geq x$. \\[\medskipamount]

$G'=G\restriction\{q\in\V{all}\mid q\text{ is reachable from }q_0\}$ \\
\+\textbf{for each} even $p\in\col(\V{all})$ \textbf{do} \\
$G_p=G'\restriction\{q\in\V{all}\mid \col(q)\geq p\}$ \\
decompose $G_p$ into SCCs \\
\+\textbf{for each} SCC~$C$ of~$G_p$ with $p\in\col(C)$ \textbf{do} \\
compute maximum cycle weight~$w$ in $C$ \\
\-\textbf{if} $w\geq x$ \textbf{then accept} \\
\-\textbf{done} \\
\textbf{done} \\
\textbf{reject}
\end{tabbing}
\vspace*{-2ex}
\caption{\label{alg:pl1-value}A polynomial-time algorithm for
deciding the value of a state in a one-player mean-payoff parity game}
\end{algorithm}
which employs as subroutines Tarjan's linear-time
algorithm~\cite{Cormen-etal09} for SCC decomposition and Karp's
polynomial-time algorithm~\cite{Karp78} for computing the
minimum\slash maximum cycle weight,
(\ie the minimum\slash maximum average weight on a cycle) in a given
strongly connected graph.

The algorithm is sound: If the algorithm accepts, then there is an even
priority~$p$ and a reachable SCC~$C$ in~$G_p$ with $p\in\col(C)$
that has maximum cycle weight $w\geq x$. 
We construct a strategy~$\sigma$ for \Pl1 with $\val^{\G}(\sigma,q_0)=w$.
Let $q\in C$ be a state with priority~$p$. Since $q$~is reachable from~$q_0$ and
$C$~is strongly connected, both $q_0$ and~$C$ lie inside $\Attr_1(\{q\})$.
Let $\sigma_q$ be the memoryless attractor strategy for~$\{q\}$. Now, since
$w$~is the maximum cycle weight in~$C$, there exists a simple cycle
$\gamma=q_1\cdots q_n q_1$ in~$C$ with cycle weight~$w$.
We construct a (memoryless) strategy $\sigma_\gamma$ on~$C$ by setting
$\sigma_\gamma(q_n)=q_1$ and
$\sigma_\gamma(q_i)=q_{i+1}$ for every $1\leq i<n$; this strategy is
extended to the whole game by combining it with an attractor
strategy for $\{q_1,\ldots,q_n\}$.
The strategies~$\sigma_q$ and~$\sigma_\gamma$ are then combined to a
strategy~$\sigma$, which is played in rounds:
in the $i$th round, \Pl1 first forces a visit
to~$\col^{-1}(p)\cap C$ by playing according to~$\sigma_q$; once
$\col^{-1}(p)\cap C$~has been reached, \Pl1 plays~$\sigma_\gamma$
for $i$~steps before proceeding to the next round.
Note that $\sigma$~fulfils the parity condition because $q$~is visited
infinitely often and all other priorities that appear infinitely often obey
$\col(q)\geq p$.
Finally, the payoff of $\rho(\sigma,q_0)$ equals the cycle weight of~$\gamma$,
\ie, $\val^{\G}(q_0)\geq\val^{\G}(\sigma,q_0)=w\geq x$. 

The algorithm is complete: Assume that $\val^{\G}(q_0)=v\geq x$ 
and let $\rho\in\Out^\G(q_0)$ be a play with $\payoff^{\G}(\rho)=v$;
such a play exists due to \cref{lem:mpp-pl1}. Consider the set
$\Inf(\rho)$ and let
$p=\min\col(\Inf(\rho))$ (which is even since $\payoff(\rho)$ is finite).
Since $\Inf(\rho)$ is strongly connected,
$\Inf(\rho)\subseteq C$ for an SCC~$C$ of~$G_p$ with $p\in\col(C)$.
Since optimal memoryless strategies exist in mean-payoff games,
there exists a simple cycle with average weight $\geq v$ in~$C$. Hence the
algorithm accepts.

Since SCC decomposition and maximum cycle weight computation both take
polynomial time, the whole algorithm runs in polynomial time.\qed
\end{proof}


It follows from \cref{thm:mpp-main,prop:mpp-pl1-ptime} that the value problem
for mean-payoff parity games is in \coNP: to~decide whether $\val^{\G}(q_0)<x$,
a nondeterministic algorithm can guess a memoryless strategy~$\tau$ for \Pl2
and check whether $\val^{\G}(\tau,q_0)<x$ in polynomial time.

\fi

\begin{corollary}\label{cor:mpp-conp}
The value problem for mean-payoff parity games is in \coNP.
\end{corollary}

\iffull
Following ideas from~\cite{CD10}, 
we prove that the value problem is not only in \coNP, but also in \NP.
The core of \cref{alg:mpp-np}
\begin{algorithm}[t]
\vspace*{.8ex}
\begin{tabbing}
\hspace*{1em}\=\hspace{1em}\=\hspace{1em}\=\hspace{1em}\=
\hspace{1em}\=\hspace{1em}\= \kill

\emph{Input:} mean-payoff parity game $\G$, state $q_0\in\V{}$,
  $x\in\bbQ$ \\[\medskipamount]

\textbf{guess} 2-trap~$T$ in~$\G$ with $q_0\in T$ \\
$\Verify(T)$\iffull \\ \else; \fi
\textbf{accept} \\[\medskipamount]

\+\textbf{procedure} $\Verify(S)$ \\
\+\textbf{if} $S\neq\emptyset$ \textbf{then} \\
$p\coloneq\min\{\col(q)\mid q\in S\}$ \\
\+\textbf{if} $p$~is even \textbf{then} \\
\textbf{guess} memoryless strategy~$\sigma_{\rmM}$ for \Pl1 in $G\restrict S$ \\
\textbf{if} $\val^{(G\restrict S,0)}(\sigma_{\rmM},q)<x$ for some $q\in S$
  \textbf{then reject} \\
$\Verify(S\setminus\Attr_1^{\G\restrict S}(\col^{-1}(p)))$ \-\\
\+\textbf{else} \\
\textbf{guess} 2-trap $T\neq\emptyset$ in
$\G\restrict (S\setminus\Attr_2^{\G\restrict S}(\col^{-1}(p)))$ \\
$\Verify(T)$;
$\Verify(S\setminus \Attr_1^{\G\restrict S}(T))$ \-\\
\textbf{end if} \-\\
\textbf{end if} \-\\
\textbf{end procedure}
\end{tabbing}
\vspace*{-2ex}
\caption{\label{alg:mpp-np}A nondeterministic algorithm for deciding
the value of a state in a mean-payoff parity game}
\end{algorithm}
is the procedure $\Verify$ that on input~$S$ checks whether the
value of all states in the game $\G\restrict S$ is at least~$x$.
If the least priority~$p$ in~$S$ is even, this is witnessed by a
strategy in the mean-payoff game $(G\restrict S,0)$ that ensures
payoff~$\geq x$ and the fact that the values of all states in
the game
$\G\restrict S\setminus\Attr_1^{\G\restrict S}(\col^{-1}(p))$
are greater than~$x$, which we can check by calling $\Verify$
recursively. If, on the other hand, the least priority~$p$ in~$S$
is odd, then $\val^{\G\restrict S}(q)\geq x$ for all $q\in S$ is
witnessed by a 2-trap~$T$ inside
$S\setminus\Attr_2^{\G\restrict S}(\col^{-1}(p))$ such that both
the values in the game $\G\restrict T$ and the values in the
game $\G\restrict S\setminus\Attr_1^{\G\restrict S}(T)$ are
bounded from below by~$x$; the latter two properties can again be
checked by calling $\Verify$ recursively.
The correctness of the algorithm relies on the following two lemmas.

\begin{lemma}\label{lemma:mpp-even}
Let $\G$ be a mean-payoff parity game with least priority~$p$
even, $T=\V{}\setminus\Attr_1(\col^{-1}(p))$, and $x\in\bbR$.
If $\val^{(G,0)}(q)\geq x$ for all $q\in\V{}$ and
$\val^{\G\restrict T}(q)\geq x$ for all $q\in T$, then
$\val^{\G}(q)\geq x$ for all $q\in\V{}$.
\end{lemma}

\begin{proof}
Assume that $\val^{(G,0)}(q)\geq x$ for all $q\in\V{}$ and
$\val^{\G\restrict T}(q)\geq x$ for all $q\in T$, and let $q^*\in\V{}$.
By \cref{thm:mpp-main}, it suffices to show that for every memoryless
strategy~$\tau$ of \Pl2 there exists a strategy~$\sigma$ of \Pl1 such that
$\payoff(\rho(\sigma,\tau,q^*))\geq x$.
Hence, assume that $\tau$~is a memoryless strategy of \Pl2 in~$\G$.
Moreover, let $\sigma_\rmM$~be a memoryless strategy for \Pl1 in~$(G,0)$ with
$\val^{(G,0)}(\sigma_\rmM,q)\geq x$ for all $q\in\V{}$, let $\sigma_T$~be a
strategy for \Pl1 in~$\G\restrict T$ with
$\val^{\G\restrict T}(\sigma_T,q)\geq x$ for all $q\in T$, and let
$\sigma_\rmA$ be a memoryless attractor strategy of \Pl1 on
$\Attr_1(\col^{-1}(p))$ that ensures to reach $\col^{-1}(p)$.
We combine these three strategies to a new strategy~$\sigma$, which is
played in rounds. In the $k$th round, the strategy behaves as follows:
\begin{enumerate}
\item while the play stays inside~$T$, play $\sigma_T$;
\item as soon as the play reaches $\Attr_1(\col^{-1}(p))$,
      switch to strategy~$\sigma_\rmA$ and play~$\sigma_\rmA$ until the
      play reaches $\col^{-1}(p)$;
\item when the play reaches $\col^{-1}(p)$, play $\sigma_M$ for exactly
      $k$~steps and proceed to the next round.
\end{enumerate}
Let $\rho\coloneqq\rho(\sigma,\tau,q^*)$. To complete the proof, we need to
show that $\payoff(\rho)\geq x$.
We distinguish whether $\rho$~visits $\Attr_1(\col^{-1}(p))$ infinitely often
or not.

In the first case, we divide~$\rho$ into $\rho=\gamma_0\gamma_1\gamma_2\cdots$
where each $\gamma_i=\gamma_i^T\gamma_i^\rmA\gamma_i^\rmM$ consists of a
part consistent with~$\sigma_T$ (thus staying inside~$T$), a part consistent
with~$\sigma_\rmA$ (thus staying in $\Attr_1(\col^{-1}(p))$),
and one that starts with a state in~$\col^{-1}(p)$ and is consistent
with~$\sigma_\rmM$.
Since $\tau$~is a memoryless strategy, there can only be $\abs{T}$~many
different~$\gamma_i^T$, and the length of each~$\gamma_i^T$ is bounded by some
constant~$k$.
Since each~$\gamma_i^\rmA$ is consistent with an attractor strategy, the length
of each~$\gamma_i^\rmA$ is bounded by $\abs{\V{}}$.
Hence, the length of $\gamma_i^\rmM$ grows continuously while the
length of~$\gamma_i^T\gamma_i^\rmA$ is bounded. Therefore,
$\liminf_{n\to\infty}\payoff_n(\rho)
=\liminf_{n\to\infty}\payoff_n(\gamma_1^\rmM\gamma_2^\rmM\cdots)$.
Since $\val^{(G,0)}(\sigma_\rmM,q)\geq x$ for all $q\in\V{}$ and
priority~$p$ is visited infinitely often, we have
$\payoff(\rho)=\liminf_{n\to\infty}\payoff_n(\rho)\geq x$.

In the second case, $\rho=\gamma\cdot\rho'$, where $\rho'$~is a play of
$\G\restrict T$ that is consistent with~$\sigma_T$. Hence,
$\payoff(\rho)=\payoff(\rho')\geq\val^{\G\restrict T}(\sigma_T,\rho'(0))
\geq x$.\qed
\end{proof}

\begin{lemma}\label{lemma:mpp-odd}
Let $\G$ be a mean-payoff parity game with least priority~$p$
odd, $T=\V{}\setminus\Attr_2(\col^{-1}(p))$, and $x\in\bbR$.
If $\val^{\G}(q)\geq x$ for some $q\in\V{}$, then $T\neq\emptyset$ and
$\val^{\G\restrict T}(q)\geq x$ for some $q\in T$.
\end{lemma}

\begin{proof}
Let $q^*\in\V{}$ be a state with $\val^{\G}(q^*)\geq 0$.
If $T=\emptyset$, then $\Attr_2(\col^{-1}(p))=\V{}$ and there is a
memoryless attractor strategy~$\tau$ for \Pl2 in~$\G$ that ensures to
visit~$\col^{-1}(p)$ infinitely often. This implies
$\val^{\G}(\tau,q^*)=-\infty$, a contradiction to
$\val^{\G}(q^*)\geq x$. Thus $T\neq\emptyset$.

Now assume that $\val^{\G\restrict T}(q)<x$ for all $q\in T$, and let $\tau$
be a (\wlg memoryless) strategy for \Pl2 in~$\G\restrict T$ that ensures
$\val^{\G\restrict T}(\tau,q)<x$ for all $q\in T$.
We~extend~$\tau$ to a strategy~$\tau'$ in~$\G$ by combining it with a
memoryless
attractor strategy for $\col^{-1}(p)$ on the states in $\V{}\setminus T$.
Let $\rho\in\Out^{\G}(\tau',q^*)$. Either $\rho$~reaches $\col^{-1}(p)$
infinitely often, in which case $\payoff^{\G}(\rho)=-\infty$, or there is a
position~$i$ from which onwards $\rho$~stays in~$T$, in which case
$\payoff^{\G}(\rho)=\payoff^{\G\restrict T}(\rho[i,\infty))\leq
\val^{\G\restrict T}(\tau,\rho(i))$. In any case,
$\val^{\G}(\tau',q^*)\leq\max_{q\in T}\val^{\G\restrict T}(\tau,q)<x$,
a~contradiction to $\val^{\G}(q^*)\geq x$.\qed
\end{proof}

Finally, \cref{alg:mpp-np} runs in polynomial time because
the value of a memoryless strategy in a mean-payoff game can be
computed in polynomial time \cite{ZP96} and because recursive
calls are limited to disjoint subarenas.

\begin{theorem}\label{thm:mpp-np}
The value problem for mean-payoff parity games is in \NP.
\end{theorem}

\begin{proof}
We claim that \cref{alg:mpp-np} is a nondeterministic polynomial-time
algorithm for the value problem. To analyse the running time, denote by
$T(n)$ the worst-case running time of the procedure $\Verify$ on a
subarena~$S$ of size~$n$. Since the value of a memoryless strategy for
\Pl1 in a mean-payoff game can be computed in polynomial time
\cite{ZP96} and attractor computations take linear time, there exists a
polynomial $f\colon\bbN\times\bbN\to\bbN$ such that the numbers
$T(n)$ satisfy the following recurrence:
\begin{align*}
 T(1) &\leq f(\size{G},\size{x}), \\
 T(n) &\leq \max_{1\leq k<n} T(k)+T(n-k)+f(\size{G},\size{x})\,.
\end{align*}
Solving this recurrence, we get that
$T(n)\leq (2n-1)f(\size{G},\size{x})$ for all $n\geq 1$,
again a polynomial. Consequently, the algorithm runs in polynomial time.

To prove the correctness of the algorithm, we need to prove that the
algorithm is both sound and complete.
We start by proving soundness: If the
algorithm accepts its input, then $\val^{\G}(q_0)\geq x$. In fact, we
prove the following stronger statement. We say that $\Verify(S)$
\emph{succeeds} if the procedure terminates without rejection
(for at least one sequence of guesses).

\begin{claim*}
Let $S\subseteq\V{}$.
If $S$~is a subarena of~$\G$ and $\Verify(S)$ does succeed,
then $\val^{\G\restrict S}(q)\geq x$ for all $q\in S$.
\end{claim*}

Assume that the claim is true and that the algorithm accepts its input.
Then there exists a 2-trap~$T$ with $q_0\in T$ such that
$\val^{\G\restrict T}(q)\geq x$ for all $q\in T$. Since $T$~is a 2-trap,
it follows that $\val^{\G}(q_0)\geq x$.

To prove the claim, we proceed by induction over the cardinality
of~$S$. If $\abs{S}=0$, the claim is trivially fulfilled. Hence,
assume that $\abs{S}>0$ and that the claim is true for all sets
$S'\subseteq\V{}$ with $\abs{S'}<\abs{S}$. Let
$p=\min\{\col(q)\mid q\in S\}$. We distinguish two cases:
\begin{enumerate}
\item The minimal priority~$p$ is even. Since
$\Verify(S)$ succeeds, there exists
a memoryless strategy~$\sigma_{\rmM}$ of \Pl1 in $\G\restrict S$ such
that $\val^{(G\restrict S,0)}(\sigma_{\rmM},q)\geq x$ for all $q\in S$,
\ie $\val^{(G\restrict S,0)}(q)\geq x$ for all $q\in S$.
Let $A=\Attr_1^{\G\restrict S}(\col^{-1}(p))$. Since $\Verify(S)$ succeeds, so
does $\Verify(S\setminus A)$. Hence, by the induction hypothesis,
$\val^{\G\restrict (S\setminus A)}(q)\geq x$ for all $q\in S\setminus A$.
By \cref{lemma:mpp-even}, these two facts imply that
$\val^{\G\restrict S}(q)\geq x$ for all $q\in S$.

\item The minimal priority~$p$ is odd. Since $\Verify(S)$ succeeds,
there exists a 2-trap $T\neq\emptyset$ in
$\G\restrict(S\setminus\Attr_2^{\G\restrict S}(\col^{-1}(p)))$ such that
both $\Verify(T)$ and $\Verify(S\setminus\Attr_1^{\G\restrict S}(T))$ succeed.
Let $A=\Attr_1^{\G\restrict S}(T))$. By the induction hypothesis, \Pl1
has a strategy~$\sigma_T$ in $\G\restrict T$ such that
$\val^{\G\restrict T}(\sigma_T,q)\geq x$ for all $q\in T$ and a
strategy~$\sigma_S$ in
$\G\restrict S\setminus A$ such that
$\val^{\G\restrict S\setminus A}(\sigma_S,q)\geq x$
for all $q\in S\setminus A$. We extend~$\sigma_T$ to a
strategy~$\sigma_A$ in $\G\restrict A$ such that
$\val^{\G\restrict A}(\sigma_A,q)\geq x$ for all $q\in A$ by
combining~$\sigma_T$ with a suitable attractor strategy.
By playing~$\sigma_S$ as long as the play
stays in $S\setminus A$ and switching
to~$\sigma_A$ as soon as the
play enters~$A$, \Pl1 can ensure that
$\val^{\G\restrict S}(q)\geq x$ for all $q\in S$.
\end{enumerate}

Finally, we prove that the algorithm is complete:
if $\val^{\G}(q_0)\geq x$, then the algorithm accepts the
input $\G,q_0,x$.
Since the set $\{q\in\V{}\mid \val^{\G}(q)\geq x\}$ is a trap for
\Pl2, it suffices to prove the following claim.

\begin{claim*}
Let $S\subseteq\V{}$. If $S$~is a subarena of~$\G$ and
$\val^{\G\restrict S}(q)\geq x$ for all $q\in S$, then
$\Verify(S)$ succeeds.
\end{claim*}

As the previous claim, we prove this claim by an induction over the
cardinality of~$S$. Clearly, $\Verify(S)$ succeeds if $\abs{S}=0$.
Hence, assume that $\abs{S}>0$ and that the claim is correct for
all sets $S'\subseteq\V{all}$ with $\abs{S'}<\abs{S}$. Moreover,
assume that $S$~is a subarena of~$\G$ such that
$\val^{\G\restrict S}(q)\geq x$ for all $q\in S$ (otherwise the
claim is trivially fulfilled). Again, we distinguish whether
$p\coloneq\min\{\col(q)\mid q\in S\}$ is even or odd.
\begin{enumerate}
 \item The minimal priority~$p$ is even. Since $\val^{\G\restrict S}(q)\geq x$
for all $q\in S$, also $\val^{(G\restrict S,0)}(q)\geq x$ for all $q\in S$,
which is witnessed by a memoryless strategy~$\sigma_{\rmM}$. Let
$A=\Attr_1^{\G\restrict S}(\col^{-1}(p))$. Since $S\setminus A$~is a
1-trap and $\val^{\G\restrict S}(q)\geq x$ for all $q\in S$, we must also
have $\val^{\G\restrict (S\setminus A)}(q)\geq x$ for all $q\in S\setminus A$.
Hence, by the induction hypothesis, $\Verify(S\setminus A)$ succeeds.
Therefore, in order to succeed, $\Verify(S)$ only needs to guess a suitable
memoryless strategy~$\sigma_{\rmM}$.

 \item The minimal priority~$p$ is odd.
Let $A\coloneq\Attr_2^{\G\restrict S}(\col^{-1}(p))$.
We claim that $\Verify(S)$ succeeds if it guesses
$T\coloneq\{q\in S\setminus A\mid\val^{\G\restrict (S\setminus A)}(q)\geq x\}$.
By \cref{lemma:mpp-odd}, the set $T$~is nonempty. Note that $T$~is a
2-trap and that
$\val^{\G\restrict T}(q)\geq x$ for all $q\in T$.
Hence, by the induction hypothesis, $\Verify(T)$~succeeds.
It remains to be shown that $\Verify(S\setminus\Attr_1^{\G\restrict S}(T))$
succeeds as well. Note that
$S\setminus\Attr_1^{\G\restrict S}(T)$~is a 1-trap, which together with
$\val^{\G\restrict S}(q)\geq x$ for all $q\in S$ implies that
$\val^{\G\restrict (S\setminus\Attr_1^{\G\restrict S}(T))}(q)\geq x$
for all $q\in S\setminus\Attr_1^{\G\restrict S}(T)$.
Hence, the induction hypothesis yields that
$\Verify(S\setminus\Attr_1^{\G\restrict S}(T))$ succeeds.\qed
\end{enumerate}
\end{proof}

\fi

\iffull
\subsection{A deterministic algorithm}
\fi
\label{sec-det}

\iffull
In this section, we present
\fi
a deterministic algorithm for computing the values of
a~mean-payoff parity game, which runs faster than all
known algorithms for solving these games. Algorithm~$\Solve$
\begin{algorithm}[t]
\vspace*{.8ex}
\begin{tabbing}
\hspace*{1em}\=\hspace{1em}\=\hspace{1em}\=\hspace{1em}\=
\hspace{1em}\=\hspace{1em}\= \kill

\textbf{Algorithm} $\Solve(\G)$ \\[\medskipamount]
\emph{Input:} mean-payoff parity game $\G=(G,\col)$ \\
\emph{Output:} $\val^\G$ \\[\medskipamount]

\textbf{if} $\V{}=\emptyset$ \textbf{then return} $\emptyset$ \\
$p\coloneq\min\{\col(q)\mid q\in\V{}\}$ \\
\+\textbf{if} $p$~is even \textbf{then} \\
  $g\coloneq\SolveMP(G,0)$ \\
  \textbf{if} $\col(q)=p$ for all $q\in\V{}$ \textbf{then return} $g$ \\
  $T\coloneq \V{}\setminus\Attr_1^{\G}(\col^{-1}(p))$;
  $f\coloneq\Solve(\G\restrict T)$ \\
  $x\coloneq\min(f(T)\cup g(\V{}))$;
  $A\coloneq\Attr_2^{\G}(f^{-1}(x)\cup g^{-1}(x))$ \\
  \textbf{return} $(\V{}\to\bbR\cup\{-\infty\}\colon q\mapsto x)\sqcup
    \Solve(\G\restrict \V{}\setminus A)$ \-\\
\+\textbf{else} \\
  $T\coloneq \V{}\setminus\Attr_2^{\G}(\col^{-1}(p))$ \\
  \textbf{if} $T=\emptyset$ \textbf{then return}
    $(\V{}\to\bbR\cup\{-\infty\}\colon q\mapsto-\infty)$ \\
  $f\coloneq\Solve(\G\restrict T)$; $x\coloneq\max f(T)$;
  $A\coloneq\Attr_1^{\G}(f^{-1}(x))$ \\
  \textbf{return} $(\V{}\to\bbR\cup\{-\infty\}\colon q\mapsto x)\sqcap
    \Solve(\G\restrict\V{}\setminus A)$ \-\\
\textbf{end if}
\end{tabbing}
\vspace*{-2ex}
\end{algorithm}
is based on the classical algorithm for solving parity
games, due to Zielonka~\cite{Zielonka98}.
The algorithm employs as a subprocedure an algorithm $\SolveMP$ for
solving mean-payoff games.
By \cite{ZP96}, such an algorithm can be implemented to run in time
$\Oh(n^3\cdot m\cdot W)$ for a game with $n$~states and $m$~edges.
We denote by $f \sqcup g$ and
$f\sqcap g$ the pointwise maximum, respectively minimum, of two
(partial) functions $f,g\colon\V{}\to\bbR\cup\{\pm\infty\}$
(where $(f\sqcup g)(q)=(f\sqcap g)(q)=f(q)$ if $g(q)$~is undefined).

The algorithm works as follows: If the least priority~$p$ in~$\calG$ is
even, the algorithm first identifies the least value of~$\G$ by computing
the values of the mean-payoff game
$(G,0)$ and (recursively) the values of the game
${\G\restrict\V{}\setminus\Attr_1(\col^{-1}(p))}$, and taking their
minimum~$x$. All states from where \Pl2 can enforce a visit to a
state with value~$x$ in one of these two games must have value~$x$
in~$\G$.
In~the remaining subarena, the values can be computed by calling
$\Solve$ recursively. If~the least priority is odd,
we can similarly compute the greatest value of~$\G$ and
proceed by recursion.
%
%

\begin{theorem}\label{thm:mpp-det}
The values of a mean-payoff parity game with $d$~priorities
can be computed in time
$\Oh(\abs{\V{}}^{d+2}\cdot \abs{E}\cdot W)$.
\end{theorem}


\begin{proof}
We claim that $\Solve$ computes, given a mean-payoff
parity game~$\G$, the function $\val^\G$ in the given time bound.
Denote by $T(n,m,d)$ the worst-case running time of the algorithm
on a game with $n$~states, $m$~edges and
$d$~priorities.
Note that, if $\G$~has only one priority, then there are no recursive
calls to $\Solve$. Since attractors can be computed in time $\Oh(n+m)$ and
the running time of $\SolveMP$ is $\Oh(n^3\cdot m\cdot W)$,
there exists a constant~$c$ such that the numbers
$T(n,m,d)$ satisfy the following recurrence:
\begin{align*}
  T(1,m,d) &\leq c, \\
  T(n,m,1) &\leq c\cdot n^3\cdot m\cdot W, \\
  T(n,m,d) &\leq T(n-1,m,d-1) + T(n-1,m,d) + c\cdot n^3\cdot m\cdot W\,.
\end{align*}
\iffull
We claim that $T(n,m,d)\leq c\cdot (n+1)^{d+2}\cdot m\cdot W\in
\Oh(n^{d+2}\cdot m\cdot W)$.
The claim is clearly true if $n=1$. Hence, assume that $n\geq 2$ and that
the claim is true for all lower values of~$n$. If $d=1$, the claim follows
from the second inequality. Otherwise,
\begin{align*}
 T(n,m,d) &\leq T(n-1,m,d-1) + T(n-1,m,d) + c\cdot n^3\cdot m\cdot W \\
&\leq c\cdot n^{d+1}\cdot m\cdot W+c\cdot n^{d+2}\cdot m\cdot W+
c\cdot n^3\cdot m\cdot W \\
&\leq c\cdot (n^{d+1}+n\cdot n^{d+1}+n^{d+1})\cdot m\cdot W \\
&\leq c\cdot ((n+1)^{d+1}+n\cdot (n+1)^{d+1})\cdot m\cdot W \\
&= c\cdot (n+1)^{d+2}\cdot m\cdot W
\end{align*}
\fi
It remains to be proved that the algorithm is correct, \ie that
$\Solve(\G)=\val^\G$.
We prove the claim by induction over the number of states. If there
are no states, the claim is trivial. Hence, assume that
$\V{}\neq\emptyset$ and that the
claim is true for all games with less than $\abs{\V{}}$~states.
Let $p\coloneq\min\{\col(q)\mid q\in\V{}\}$. We only consider the case that
$p$~is even. If $p$~is odd, the proof is similar, but relies on
\cref{lemma:mpp-odd} instead of \cref{lemma:mpp-even}.

Let $T$, $f$, $g$, $x$ and~$A$ be defined as in the corresponding case
of the algorithm, and let $f^*=\Solve(\G)$.
If $\col(\V{})=\{p\}$, then $f^*=g=\val^{(G,0)}=
\val^{\G}$, and the claim is fulfilled. Otherwise,
by the definition of~$x$ and applying the induction hypothesis to
the game $\G\restrict T$, we have $\val^{(G,0)}(q)\geq x$ for
all $q\in\V{}$ and $\val^{\G\restrict T}(q)=f(q)\geq x$ for all $q\in T$.
Hence, \cref{lemma:mpp-even} yields that $\val^{\G}(q)\geq x$
for all $q\in\V{}$. On the other hand, from any state $q\in A$ \Pl2 can
play an attractor strategy to $f^{-1}(x)\cup g^{-1}(x)$, followed by an
optimal strategy in the game $\G\restrict T$, respectively in the
mean-payoff game $(G,0)$, which ensures that \Pl1's payoff
does not exceed~$x$.
Hence, $\val^{\G}(q)=x=f^*(q)$ for all $q\in A$.

Now, let $q\in\V{}\setminus A$. We already know that $\val^{\G}(q)
\geq x$. Moreover, since $\V{}\setminus A$ is a 2-trap and applying
the induction hypothesis to the game $\G\restrict\V{}\setminus A$, we have
$\val^{\G}(q)\geq\val^{\G\restrict\V{}\setminus A}(q)=
\Solve(\G\restrict\V{}\setminus A)(q)$.
Hence, $\val^{\G}(q)\geq f^*(q)$. To see that $\val^{\G}(q)\leq f^*(q)$,
consider the strategy~$\tau$ of \Pl2 that mimics an optimal strategy
in $\G\restrict\V{}\setminus A$ as long as the play stays in
$\V{}\setminus A$
and switches to an optimal strategy in~$\G$ as soon as the
play reaches~$A$. We have $\val^{\G}(\tau,q)\leq
\max\{\val^{\G\restrict\V{}\setminus A}(q),x\}=f^*(q)$.\qed
\end{proof}


Algorithm~$\Solve$ is faster and conceptually simpler than the original
algorithm proposed for solving mean-payoff parity games~\cite{CHJ05}.
Compared to the recent algorithm proposed by Chatterjee and Doyen~\cite{CD10},
which uses a reduction to energy parity games and runs in time
$\Oh(\abs{\V{}}^{d+4}\cdot \abs{E}\cdot d\cdot W)$, our algorithm has three
main advantages: 1.~it is faster; 2.~it operates directly on mean-payoff
parity games, and 3.~it is more flexible since it computes the values
exactly instead of just comparing them to an integer threshold.

\section{Mean-penalty parity games}
\label{sec-mpep}

\iffull
In this second part of the paper,
\fi
we define multi-strategies and \emph{mean-penalty parity games}.
We~reduce these games to mean-payoff parity games, show that
their value problem is in $\NP\cap\coNP$, and propose
a deterministic algorithm for computing the values, which runs
in pseudo-polynomial time if the number of priorities is bounded.

\iffull
\subsection{Definitions}
\fi

Syntactically, a~\emph{mean-penalty parity game} is a mean-payoff
parity game with non-negative weights, \ie a tuple $\G=(G,\col)$, where
$G=(\V1,\V2,E,\weight)$~is a weighted game graph with $\weight\colon E
\to \bbR^{\geq 0}$ (or $\weight\colon E\to\bbN$ for algorithmic
purposes), and $\col\colon \V{all}\to\bbN$ is a
priority function assigning a priority to every state.
As~for mean-payoff parity games, a~play~$\rho$ is
parity-winning if the minimal priority occurring infinitely often
($\min\{\col(q)\mid q\in\Inf(\rho)\}$) is even.

Since we are interested in controller synthesis, we define
multi-strategies only for \Pl1 (who represents the system).
Formally, a~\emph{multi-strategy} (for \Pl1) in~$\G$ is a function
$\sigma\colon \V{all}^* \V{1} \to\pow(\V{all})\setminus\{\emptyset\}$
such that $\sigma(\gamma q)\subseteq qE$ for all $\gamma\in\V{}^*$
and $q\in\V{1}$.
A~play~$\rho$ of~$\G$ is \emph{consistent} with a multi-strategy~$\sigma$
if $\rho(k+1)\in\sigma(\rho[0,k])$ for all $k\in\bbN$ with $\rho(k)\in\V{1}$,
and we denote by $\Out^{\G}(\sigma,q_0)$ the set of all
plays~$\rho$ of~$\G$ that are consistent with~$\sigma$ and start in
$\rho(0)=q_0$.

Note that, unlike for deterministic strategies, there is, in general,
no unique play consistent with a multi-strategy~$\sigma$ for \Pl1 and a
(deterministic) strategy~$\tau$ for \Pl2 from a given initial state. Finally,
note that every deterministic strategy can be viewed as a multi-strategy.

Let $\G$ be a mean-penalty parity game, and let $\sigma$ be a multi-strategy.
We~inductively define $\pnlty^{\G}_{\sigma}(\gamma)$ (the
\emph{total penalty} of~$\gamma$ \wrt~$\sigma$) for all $\gamma\in\V{}^*$
by setting $\pnlty^{\G}_{\sigma}(\epsilon)=0$ as well as
$\pnlty^{\G}_{\sigma}(\gamma q)=\pnlty^{\G}_{\sigma}(\gamma)$ if
$q\in\V2$ and
\begin{equation*}
\pnlty^{\G}_{\sigma}(\gamma q)=
 \pnlty^{\G}_{\sigma}(\gamma)
 +\sum_{\mathmakebox[0.8cm][c]{q'\in qE\setminus\sigma(\gamma q)}}
 \weight(q,q')
\end{equation*}
if $q\in\V1$.
Hence, $\pnlty^{\G}_{\sigma}(\gamma)$ is the total weight of transitions
blocked by~$\sigma$ along~$\gamma$.
The \emph{mean penalty} of an infinite play~$\rho$ is then defined as
the average penalty that is incurred along this play in the limit,
\ie
\begin{equation*}
\pnlty^{\G}_{\sigma}(\rho)=
  \begin{cases}\displaystyle
  \limsup_{n\to\infty}\tfrac{1}{n}\pnlty^{\G}_{\sigma}(\rho[0,n)) &
    \text{if $\rho$~is parity-winning,} \\
  \infty & \text{otherwise.}
  \end{cases}
\end{equation*}
The mean penalty of a strategy~$\sigma$ from a given initial state~$q_0$
is defined as the supremum over the mean penalties of all plays that
are consistent with~$\sigma$, \ie
\begin{equation*}
\pnlty^{\G}(\sigma,q_0)=
\sup\{\pnlty^{\G}_{\sigma}(\rho)\mid\rho\in\Out^{\G}(\sigma,q_0)\}.
\end{equation*}
The \emph{value} of a state~$q_0$ in a mean-penalty parity game
$\G$ is the least mean penalty that a multi-strategy of \Pl1 can achieve, \ie
$\val^{\G}(q_0)=\inf_{\sigma}\pnlty^{\G}(\sigma,q_0)$, where $\sigma$~ranges
over all multi-strategies of \Pl1. A~multi-strategy~$\sigma$ is
called \emph{optimal} if
$\pnlty^\G(\sigma,q_0)=\val^\G(q_0)$ for all $q_0\in Q$.

Finally, the \emph{value problem for mean-penalty parity games} is the
following decision problem:
Given a mean-penalty parity game $\G=(G,\col)$,
an initial state $q_0\in\V{all}$, and a
number $x\in\bbQ$, decide whether $\val^{\G}(q_0)\leq x$.

\begin{figure}
\begin{floatrow}
\ffigbox[.3\textwidth]{%
  \centering
  \begin{tikzpicture}[x=1.5cm,y=1.2cm]
    \draw (0,0) node[pl1,label={above:$1$}] (a) {} node {$q_1$};
    \draw (0,-2) node[pl2,label={left:$0$}] (b) {} node {$q_2$};
    \draw[->] (a) to [bend right] node[midway,left] {$1$} (b);
    \draw[->] (a) to [loop,out=140,in=-140,looseness=4] node[midway,left] {$2$} (a);
    \draw[->] (b) to [bend right] (a);
\end{tikzpicture}
}{\caption{\label{fig:mpep}A mean-penalty parity game}}
\ffigbox[.6\textwidth]{%
  \centering
  \begin{tikzpicture}[x=1.2cm,y=1.2cm]
    \draw (0,0) node[pl1,label={above:$1$}] (a) {} node {$q_1$};
    \draw (0,-2) node[pl2,label={left:$0$}] (b) {} node {$q_2$};
    \draw (-1.5,-1) node[pl2,fill=black!10!white] 
      (a1) {$(q_1,\{q_1\})$};
    \draw (0,-1) node[pl2,fill=black!10!white] 
      (a2) {$(q_1,\{q_1,q_2\})$};
    \draw (1.5,-1) node[pl2,fill=black!10!white] 
      (a3) {$(q_1,\{q_2\})$};
    \draw (3,-1) node[pl2,fill=black!10!white] 
      (b1) {$(q_2,\{q_1\})$};
    \draw[->] (a) -- (a1)
      node[midway,above] {$-2\,\,\,\,$};
    \draw[->] (a.-70) .. controls +(-70:2mm) and +(90:2mm) .. (a2.60)
      node[midway,right] {$0$};
    \draw[->] (a) -- (a3) 
      node[midway,above] {$\,\,-4$};
    \draw[->] (a1.90) .. controls +(90:10mm) and +(180:10mm) .. (a);
    \draw[->] (a2.120) .. controls +(90:2mm) and +(-110:2mm) .. (a.-110);
    \draw[->] (a3) -- (b);
    \draw[->] (a2) -- (b);
    \draw[->,rounded corners=4mm] (b) -| (b1);
    \draw[->,rounded corners=4mm] (b1) |- (a);
\end{tikzpicture}
}{\caption{\label{fig:mppofmpep}The corresponding mean-payoff parity game}}
\end{floatrow}
\end{figure}

\begin{example}\label{ex:mpep}
Fig.~\ref{fig:mpep} represents a mean-penalty parity game. Note that
weights of transitions out of \Pl2 states are not indicated as they are
irrelevant for the mean penalty.
In this game, \Pl1 (controlling circle states) has to regularly \emph{block}
the self-loop if~she wants to enforce infinitely many visits to the
state with priority~$0$.
This comes with a penalty of~$2$. However, the multi-strategy in which she
blocks no transition can be played safely for an arbitrary number of times. Hence
\Pl1 can win with mean-penalty~$0$ (but infinite memory), by blocking
the self-loop once every $k$~moves, where $k$~grows with the number of
visits to~$q_2$.
\end{example}

\iffull
\subsection{Strategy complexity}
\fi
\label{sec:mean-penalty_mean-payoff}

In order to solve mean-penalty games, we reduce them to
mean-payoff parity games.
We~construct from a given mean-penalty parity game~$\G$ an exponential-size
mean-payoff parity game~$\G'$, similar to~\cite{BDMR09} but with an added
priority function.
%
Formally,
for a mean-penalty parity game $\G=(G,\col)$ with game graph
$G=(\V1,\V2,E,\weight)$, the game graph $G'=(\V1',\V2',E',\weight')$
of the corresponding mean-payoff parity game~$\G'$ is defined as follows:
\begin{itemize}
\item $\V1'=\V1$ and $\V2'=\V2\cup\bar{\V{all}}$, where
$\bar{\V{}}\coloneq\{(q,F)\mid q\in\V{all},\ \emptyset\neq F\subseteq qE\}$;
\item $E'$ is the (disjoint) union of three kinds of transitions:
	\begin{enumerate}[(1)]
	\item transitions of the form $(q,(q,F))$ for each $q\in\V1$ and
	$\emptyset\neq F\subseteq qE$,
	\item transitions of the form $(q,(q,\{q'\}))$ for each $q\in\V2$
	and $q'\in qE$,
	\item transitions of the form $((q,F),q')$ for each $q'\in F$;
	\end{enumerate}
\item the weight function $\weight'$ assigns~$0$ to transitions of type~(2)
  and~(3), but
  $\weight'(q,(q,F))=-2\sum_{q'\in qE\setminus F}\weight(q,q')$
to transitions of type~(1).
\end{itemize}
Finally, the priority function~$\col'$
of~$\G'$ coincides with~$\col$ on~$\V{all}$ and assigns priority
$M\coloneq\max\{\col(q) \mid q\in \V{all}\}$
to all states in~$\bar{\V{all}}$.

\begin{example}
\cref{fig:mppofmpep} depicts the
mean-payoff parity game obtained from the mean-penalty parity game
from \cref{ex:mpep}, depicted in \cref{fig:mpep}.
\end{example}

\noindent
The correspondence between~$\G$ and~$\G'$ is expressed in the following lemma.

\begin{lemma}
\label{lem:value_equivalence}
Let $\G$ be a mean-penalty parity game,
$\G'$~the corresponding mean-payoff parity game,
and $q_0\in Q$.
\begin{enumerate}
\item For every multi-strategy~$\sigma$ in~$\G$ there exists a
strategy~$\sigma'$ for \Pl1 in~$\G'$ such that
$\val(\sigma',q_0)\geq-\pnlty(\sigma,q_0)$.

\item For every strategy~$\sigma'$ for \Pl1 in~$\G'$ there exists
a multi-strategy~$\sigma$ in~$\G$ such that
$\pnlty(\sigma,q_0)\leq-\val(\sigma',q_0)$.

\item $\val^{\G'}(q_0)=-\val^{\G}(q_0)$.
\end{enumerate}
\end{lemma}

\iffull
\begin{proof}
Clearly, 3.~is implied by 1.~and 2., and we only need to prove the
first two statements. To prove 1., let $\sigma$~be a multi-strategy
in~$\G$. For a play prefix $\gamma=q_0(q_0,F_0)\cdots q_n(q_n,F_n)$
in~$\G'$, let $\tilde{\gamma}\coloneq q_0\cdots q_n$ be the
corresponding play prefix in~$\G$. We set
$\sigma'(\gamma q)=(q,F)$ if $q\in\V{1}$ and
$\sigma(\tilde{\gamma} q)=F$.
Clearly, for each
$\rho'\in\Out(\sigma',q_0)$ there exists a play
$\rho\in\Out(\sigma,q_0)$ with
$-\pnlty_\sigma(\rho)=\payoff(\rho')$
(namely $\rho(i)=\rho'(2i)$ for all ${i\in\bbN}$). Hence,
\begin{align*}
\val^{\G'}(\sigma',q_0)
&=\inf\{\payoff(\rho')\mid\rho'\in\Out(\sigma',q_0)\} \\
&\geq\inf\{-\pnlty_\sigma(\rho)\mid\rho\in\Out(\sigma,q_0)\} \\
&=-\sup\{\pnlty_\sigma(\rho)\mid\rho\in\Out(\sigma,q_0)\} \\
&=-\pnlty(\sigma,q_0)\,.
\end{align*}

To prove 2., let $\sigma'$ be a strategy for \Pl1 in~$\G'$. For a play prefix
$\gamma=q_0\cdots q_n$ in~$\G$, we inductively define the corresponding play
prefix $\tilde{\gamma}$ in~$\G'$ by setting
$\tilde{q}=q$ and $\tilde{\gamma q}=
\tilde{\gamma}\cdot\sigma'(\tilde{\gamma})\cdot q$. We set
$\sigma(\gamma)=F$ if $\sigma'(\tilde{\gamma})=(q,F)$.
For each $\rho\in\Out(\sigma,q_0)$ there exists a play
$\rho'\in\Out(\sigma',q_0)$ with $\pnlty_\sigma(\rho)=-\payoff(\rho')$,
namely the play~$\rho'$ defined by $\rho'(2i)=\rho(i)$ and
\begin{align*}
\rho'(2i+1)=\begin{cases}
	(\rho(i),\sigma(\rho[0,i])) & \text{if $\rho(i)\in\V1$,} \\
	(\rho(i),\{\rho(i+1)\})	&\text{if $\rho(i)\in\V2$,}
\end{cases}
\end{align*}
for all $i\in\bbN$. Hence,
\begin{align*}
\pnlty(\sigma,q_0)
&=\sup\{\pnlty_\sigma(\rho)\mid\rho\in\Out(\sigma,q_0)\} \\
&\leq\sup\{-\payoff(\rho')\mid\rho'\in\Out(\sigma',q_0)\} \\
&=-\inf\{\payoff(\rho')\mid\rho'\in\Out(\sigma',q_0)\} \\
&=-\val^{\G'}(\sigma',q_0)\,. \tag*{\qed}
\end{align*}
\end{proof}
\fi

It~follows from \cref{thm:mpp-main,lem:value_equivalence} that
every mean-penalty parity game admits an optimal multi-strategy.

\begin{corollary}
In every mean-penalty parity game, \Pl1 has an optimal
multi-strategy.
\end{corollary}

We now show that \Pl2 has a memoryless optimal strategy of a special kind
in the mean-payoff parity game derived from a mean-penalty parity game.
This puts the value problem for mean-penalty parity games into \coNP, and is
also a crucial point in the proof of Lemma~\ref{lemma:value_equivalence_2} below.

\begin{lemma}
\label{lem:total_order}
Let $\G$ be a mean-penalty parity game and $\G'$ the corresponding
mean-payoff parity game.  Then in~$\G'$ there is a memoryless optimal
strategy~$\tau'$ for \Pl2 such that for every $q\in\V{all}$ there
exists a total order $\leq_q$ on the set~$qE$ with
$\tau'((q,F))=\min_{\leq_q}F$ for every state $(q,F)\in\bar{\V{all}}$.
\end{lemma}

\iffull
\begin{proof}
\else
\begin{proof}[Sketch]
\fi
Let $\tau$~be a memoryless optimal strategy for \Pl2 in~$\G'$. For a
state~$q$, we consider the set~$qE$ and order it in the following way.
We inductively define $F_1=qE$, $q_i=\tau((q,F_i))$ and
$F_{i+1}=F_i\setminus\{q_i\}$ for every $1\leq i\leq\abs{qE}$.  Note
that $\{q_1,\ldots,q_{\abs{qE}}\}=qE$.  We set
$q_1\leq_q q_2\leq_q\cdots\leq_q q_{\abs{qE}}$ and define a new
memoryless strategy~$\tau'$ for \Pl2 in ~$\G'$ by
$\tau'((q,F))=\min_{\leq_q}F$ for $(q,F)\in\bar{\V{all}}$
and $\tau'(q)=\tau(q)$ for all $q\in\V2$.
\iffull
To prove the lemma, we have to show that $\tau'$~is at least as good
as~$\tau$ and thus optimal.

Let $q_0\in\V{all}$ and $\rho'\in\Out(\tau',q_0)$.  We construct
a play $\rho\in\Out(\tau,q_0)$ with
$\payoff(\rho)\geq\payoff(\rho')$ in the following way. For every
position~$i$ with $\rho'(i)=(q,F')$, let $F=\{q'\in qE\mid
\tau'((q,F'))\leq_q q'\}$ (then $\tau((q,F))=\tau'((q,F'))$ by the
definition of~$\tau'$) and set $\rho(i)=(q,F)$. For every other
position~$i$, let $\rho(i)=\rho'(i)$.  Note that
$\rho\in\Out(\tau,q_0)$ and
$\min\col(\Inf(\rho))=\min\col(\Inf(\rho'))$.  Moreover,
we have $F'\subseteq F$ and therefore
$\weight'(q,(q,F'))\leq\weight'(q,(q,F))$ whenever
$\rho'(i)=(q,F')$ and $\rho(i)=(q,F)$
(because weights in~$\G$ are nonnegative).
Hence, $\payoff(\rho)\geq\payoff(\rho')$.
Since $\rho'$~was chosen arbitrarily, it follows that
\begin{align*}
\val(\tau,q_0)
&=\sup\{\payoff(\rho)\mid\rho\in\Out(\tau,q_0)\} \\
&\geq\sup\{\payoff(\rho')\mid\rho'\in\Out(\tau',q_0)\} \\
&=\val(\tau',q_0)\,.
\end{align*}
Hence, $\tau'$~is optimal.\qed
\fi
\end{proof}

\iffull
\subsection{Computational complexity}
\fi

In order to put the value problem for mean-penalty parity games into
${\NP\cap\coNP}$, we propose a more sophisticated reduction
from mean-penalty parity games to mean-payoff parity games, which
results in a polynomial-size mean-payoff parity game. Intuitively,
in a state $q\in\V{1}$ we ask \Pl1 \emph{consecutively} for each
outgoing transition whether he wants to block that transition.
If he allows a transition, then \Pl2 has to decide whether she wishes
to explore this transition. Finally, after all transitions have been
processed in this way, the play proceeds along the \emph{last}
transition that \Pl2 has desired to explore.

Formally, let us fix a mean-penalty parity game $\G=(G,\col)$ with
game graph $G=(\V{1},\V{2},E,\weight)$, and denote by
$k\coloneq\max\{\abs{qE}\mid q\in\V{}\}$ the maximal out-degree of a state.
Then the polynomial-size mean-payoff
parity game~$\G''$ has vertices of the form $q$ and~$(q,a,i,m)$, where
$q\in\V{}$, $a\in\{\select,\allow,\block\}$, $i\in\{1,\dots,k+1\}$ and
$m\in\{0,\ldots,k\}$;
vertices of the form $q$ and $(q,\select,i,m)$ belong to \Pl1, while
vertices of the form $(q,\allow,i,m)$ or $(q,\block,i,m)$ belong
to \Pl2.
To describe the transition structure of~$\G$, let $q\in\V{}$ and
assume that $qE=\{q_1,\ldots,q_k\}$ (a state may occur more than once in
this list). Then the following transitions
originate in a state of the form $q$ or~$(q,a,i,m)$:
\begin{enumerate}
\item a transition from $q$ to~$(q,\select,1,0)$ with weight~$0$,
\item for all $1\leq i\leq k$ and $0\leq m\leq k$ a transition from
 $(q,\select,i,m)$ to $(q,\allow,i,m)$ with weight~$0$,
\item if $q\in\V{1}$ then for all $1\leq i\leq k$ and $0\leq m\leq k$
 a transition from $(q,\select,i,m)$ to $(q,\block,i,m)$ with
 weight~$0$, \emph{except} if $i=k$ and $m=0$;
\item for all $0\leq m\leq k$ a transition from $(q,\select,k+1,m)$ to~$q_m$
 with weight~$0$ (where $q_0$~can be chosen arbitrarily),
\item for all $1\leq i\leq k$ and $0\leq m\leq k$ a transition from
 $(q,\allow,i,m)$ to $(q,\select,i+1,i)$ with weight~$0$,
\item for all $1\leq i\leq k$ and $1\leq m\leq k$ a transition from
 $(q,\allow,i,m)$ to $(q,\select,i+1,m)$ with weight~$0$,
\item for all $1\leq i\leq k$ and $0\leq m\leq k$ a transition from
 $(q,\block,i,m)$ to $(q,\select,i+1,m)$ with weight
 $-2(k+1)\cdot\weight(q,q_i)$.
\end{enumerate}
Finally, the priority of a state $q\in\V{}$ equals the
priority of the same state in~$\G$, whereas all states of the
form $(q,a,i,m)$ have priority $M=\max\{\col(q)\mid q\in\V{}\}$.

\begin{example}
For the game of \cref{fig:mpep}, this transformation would yield the game
depicted in \cref{fig:mpeptopolympp}.
\begin{figure}
\centering
\begin{tikzpicture}[x=2cm,y=1.2cm]
\draw[rounded corners=3mm,dashed] (1,2.3) -| (5.5,-1.8) -| (-0.5,-1) |- (1,2.3);
    \draw (0,0) node[pl1,label={left:$1$}] (q0) {} node {$q_1$};
    \draw (1,0) node[pl1big,label={below:$1$}] (q0c10) {} node {\four{q_1}c10};
    \draw (2,.9) node[pl2big,label={below:$1$}] (q0a10) {} node {\four{q_1}a10};
    \draw (2,-.9) node[pl2big,label={below:$1$}] (q0b10) {} node {\four{q_1}b10};
    \draw (3,.9) node[pl1big,label={below:$1$}] (q0c21) {} node {\four{q_1}c21};
    \draw (3,-.9) node[pl1big,label={below:$1$}] (q0c20) {} node {\four{q_1}c20};
    \draw (4,-.9) node[pl2big,label={below:$1$}] (q0a20) {} node {\four{q_1}a20};
    \draw (4,.3) node[pl2big,label={below:$1$}] (q0a21) {} node {\four{q_1}a21};
    \draw (4,1.5) node[pl2big,label={below:$1$}] (q0b21) {} node {\four{q_1}b21};
    \draw (5,.9) node[pl1big,label={below:$1$}] (q0c31) {} node {\four{q_1}c31};
    \draw (5,-.9) node[pl1big,label={above:$1$}] (q0c32) {} node {\four{q_1}c32};
\begin{scope}[yshift=-8mm]
\draw[rounded corners=3mm,dashed] (1,-1.4) -| (5.5,-2.8) -| (-0.5,-2.2) |- (1,-1.4);
    \draw (5,-2) node[pl1,label={below:$0$}] (q1) {} node {$q_2$};
    \draw (4,-2) node[pl1big,label={below:$1$}] (q1c10) {} node {\four{q_2}c10};
    \draw (3,-2) node[pl2big,label={below:$1$}] (q1a10) {} node {\four{q_2}a10};
    \draw (2,-2) node[pl1big,label={below:$1$}] (q1c21) {} node {\four{q_2}c21};
    \draw (1,-2) node[pl2big,label={below:$1$}] (q1a21) {} node {\four{q_2}a21};
    \draw (0,-2) node[pl1big,label={below:$1$}] (q1c31) {} node {\four{q_2}c31};
\end{scope}
    \draw[->] (q0) -- (q0c10);
    \draw[->] (q0c10) -- (q0a10);
    \draw[->] (q0c10) -- (q0b10);
    \draw[->] (q0a10) -- (q0c21);
    \draw[->] (q0b10) -- node[above] {$-12$} (q0c20);
    \draw[->] (q0c21) -- (q0a21);
    \draw[->] (q0c21) -- (q0b21);
    \draw[->] (q0c20) -- (q0a20);
    \draw[->] (q0a21) -- (q0c31);
    \draw[->] (q0a21) -- (q0c32);
    \draw[->] (q0b21) -- node[above] {$-6$} (q0c31);
    \draw[->] (q0a20) -- (q0c32);
    \draw[->] (q0c32) -- (q1);
    \draw[->,rounded corners=2mm] (q0c31) --  +(0,1.2)
      node[coordinate] (z) {} -| (q0);
    \draw[->] (q1) -- (q1c10);
    \draw[->] (q1c10) -- (q1a10);
    \draw[->] (q1a10) -- (q1c21);
    \draw[->] (q1c21) -- (q1a21);
    \draw[->] (q1a21) -- (q1c31);
    \draw[->] (q1c31) -- (q0);
\end{tikzpicture}
\caption{The game~$\calG''$ associated with the game~$\calG$  of
  \cref{fig:mpep}}
\label{fig:mpeptopolympp}
\end{figure}
In this picture, $\rma$, $\rmb$
and~$\rmc$ stand for \emph{allow}, \emph{block} and \emph{choose},
respectively; zero weights are omitted.
\end{example}

It is easy to see that the game~$\calG''$ has polynomial size and
can, in fact, be constructed in polynomial time from the given
mean-penalty parity game~$\calG$. The following lemma relates
the game~$\calG''$ to the mean-payoff parity game~$\calG'$ of
exponential size constructed
\iffull
in \cref{sec:mean-penalty_mean-payoff}
\fi
and to the original game~$\calG$.

\begin{lemma}
\label{lemma:value_equivalence_2}
Let $\G$ be a mean-penalty parity game,
$\G'$~the corresponding mean-payoff parity game of exponential size,
$\G''$~the corresponding mean-payoff parity game of polynomial size,
and $q_0\in Q$.
\begin{enumerate}
\item For every multi strategy~$\sigma$ in~$\G$ there exists a
strategy~$\sigma'$ for \Pl1 in~$\G''$ such that
$\val(\sigma',q_0)\geq-\pnlty(\sigma,q_0)$.

\item For every strategy~$\tau$ for \Pl2 in~$\G'$ there exists
a strategy~$\tau'$ for \Pl2 in~$\G''$ such that
$\val(\tau',q_0)\leq\val(\tau,q_0)$.

\item $\val^{\G''}(q_0)=-\val^{\G}(q_0)$.
\end{enumerate}
\end{lemma}

\iffull
\begin{proof}
To prove 1.,
let $\sigma$~be a multi-strategy in~$\G$. For any play prefix
$\gamma$ in~$\G''$, let $\tilde{\gamma}$ be the projection to states
in~$\G$ (\ie all states of the form $(q,a,i,m)$ are omitted).
Assuming that $q_1,\ldots,q_k$ is the enumeration of~$qE$ used in the
definition of~$\G''$,
we set $\sigma'(\gamma\cdot(q,\select,i,m))=(q,\allow,i,m)$ if (and only if)
either $q\in\V{1}$ and $q_i\in\sigma(\tilde{\gamma})$ or
$q\in\V{2}$.
It is easy to see that for each
$\rho'\in\Out(\sigma',q_0)$ there exists a play
$\rho\in\Out(\sigma,q_0)$ with
$-\pnlty_\sigma(\rho)=\payoff(\rho')$. Hence,
\begin{align*}
\val(\sigma',q_0)
&=\inf\{\payoff(\rho')\mid\rho'\in\Out(\sigma',q_0)\} \\
&\geq\inf\{-\pnlty_\sigma(\rho)\mid\rho\in\Out(\sigma,q_0)\} \\
&=-\sup\{\pnlty_\sigma(\rho)\mid\rho\in\Out(\sigma,q_0)\} \\
&=-\pnlty(\sigma,q_0)\,.
\end{align*}

To prove 2., let $\tau$ be a strategy for \Pl2 in~$\G'$. By
\cref{lem:total_order}, there exists a memoryless strategy~$\tau^*$ for
\Pl2 in~$\G'$ such that $\val(\tau^*,q_0)\leq\val(\tau,q_0)$
and for all $q\in\V{}$ there exists a total order~$\leq_q$
on~$qE$ with $\tau^*((q,F))=\min_{\leq_q} F$ for all $(q,F)\in\bar{\V{}}$.
We define a memoryless strategy~$\tau'$ for \Pl2 in~$\G''$ as follows:
Assume that $q_1,\ldots,q_k$ is the enumeration of~$qE$ used in the
definition of~$\G''$. Then we set
$\tau'((q,\allow,i,m))=(q,\select,i+1,i)$ if (and only if)
one of the following three conditions is fulfilled:
1.~$m=0$, or 2.~$q\in\V{1}$ and $q_i\leq_q q_m$, or
3.~$q\in\V{2}$ and $\tau^*(q)=(q,\{q_i\})$.
Now it is easy to see that for each $\rho'\in\Out(\tau',q_0)$
there exists a play $\rho\in\Out(\tau^*,q_0)$ with
$\payoff(\rho)=\payoff(\rho')$. Hence,
\begin{align*}
\val(\tau',q_0)
&=\sup\{\payoff(\rho')\mid\rho'\in\Out(\tau',q_0)\} \\
&\leq\sup\{\payoff(\rho)\mid\rho\in\Out(\tau^*,q_0)\} \\
&=\val(\tau^*,q_0) \\
&\leq\val(\tau,q_0)\,.
\end{align*}

Finally, we prove~3. It follows from 1.\ that
$\val^{\G''}(q_0)\geq-\val^{\G}(q_0)$, and it follows
from 2.\ that $\val^{\G''}(q_0)\leq\val^{\G'}(q_0)$.
But $\val^{\G'}(q_0)=-\val^{\G}(q_0)$ by
\cref{lem:value_equivalence}, and therefore
$\val^{\G''}(q_0)=-\val^{\G}(q_0)$.\qed
\end{proof}
\fi

Since the mean-payoff game~$\G''$ can be computed from~$\G$ in polynomial
time, we obtain a polynomial-time many-one reduction from the
value problem for mean-penalty parity games
to the value problem for mean-payoff parity games.
By \cref{cor:mpp-conp,thm:mpp-np}, 
the latter problem belongs to ${\NP\cap\coNP}$.

\begin{theorem}
\label{thm:value-penalty-NPcoNP}
The value problem for mean-penalty parity games belongs to
${\NP\cap\coNP}$.
\end{theorem}

\iffull
\subsection{A deterministic algorithm}
\fi

Naturally, we can use the polynomial translation from mean-penalty parity
games to mean-payoff parity games to solve mean-penalty parity games
deterministically. Note that the mean-payoff parity game~$\G''$
derived from a mean-penalty parity game has $\Oh(\abs{\V{}}\cdot k^2)$
states and $\Oh(\abs{\V{}}\cdot k^2)$ edges, where $k$~is the maximum
out-degree of a state in~$\G$; the number of priorities
remains constant.
Moreover, if weights are given in integers and $W$~is the highest absolute
weight in~$\G$, then the highest absolute weight in~$\G''$ is $\Oh(k\cdot W)$.
Using \cref{thm:mpp-det}, we thus obtain a deterministic algorithm for
solving mean-penalty parity games that runs in time
$\Oh(\abs{\V{}}^{d+3}\cdot k^{2d+7}\cdot W)$.
If $k$~is a constant, the running time is
$\Oh(\abs{\V{}}^{d+3}\cdot W)$, which is acceptable. In the
general case however, the best upper bound on~$k$ is the number of
states, and we get an algorithm that runs in time
$\Oh(\abs{\V{}}^{3d+10}\cdot W)$. Even if the numbers of priorities
is small, this running time would not be acceptable in practical
applications.

The goal of this section is to show
that we can do better; namely we will give an algorithm that runs
in time $\Oh(\abs{\V{}}^{d+3}\cdot\abs{E}\cdot W)$, independently of
the maximum out-degree. The~idea is as follows: we~use
Algorithm~$\Solve$ on the mean-payoff parity game~$\G'$
of exponential size, but we show that we can run it \emph{on~$\G$}, \ie,
by handling the extra states of~$\G'$ symbolically during the computation.
As a first step, we adapt the pseudo-polynomial algorithm
by Zwick and Paterson~\cite{ZP96} to compute
the values of a mean-penalty parity game with a trivial parity
objective.

\begin{lemma}\label{lemma-ZP}
The values of a mean-penalty parity game with priority
function $\col\equiv 0$ can be computed in time
$\Oh(\abs{\V{all}}^4\cdot \abs E\cdot W)$.
\end{lemma}

\iffull
\begin{proof}
Let $\G=(G,\col)$ with
$G=(Q_1,Q_2,E,\weight)$, and $\G'=(G',\col')$ with
$G'=(Q'_1,Q'_2,E',\weight')$.
For a state~$q\in \V{all}'$, we~let $v_0(q)=0$, and
for~$k>0$,  we~define
\[
v_k(q) = \begin{cases}
  \displaystyle
  \max_{q'\in qE'} \weight'(q,q') + v_{k-1}(q') &\text{if $q\in \V1'$,}  \\
  \displaystyle
  \min_{q'\in qE'} \weight'(q,q') + v_{k-1}(q') &\text{if $q\in \V2'$.}
	 \end{cases}
\]
If $q\in\V{all}$, then the definition of~$\G'$ yields that
\[
v_k(q) = \begin{cases}\displaystyle
  \max_{F\subseteq qE} \weight'(q,(q,F))  +
\min_{q'\in F} v_{k-2}(q') & \text{if $q\in\V{1}$,} \\
  \displaystyle
  \min_{q'\in qE} v_{k-2}(q') & \text{if $q\in\V{2}$,}
  \end{cases}
\]
In the first case, a~na\"ive computation would require the examination
of an exponential
number of transitions. In order to avoid this blow-up, we~use the same
idea as in the proof of Lemma~\ref{lem:total_order}:
Let $qE=\{q_1,\dots,q_r\}$ be
sorted in such a way that $i\leq j$ implies $v_{k-2}(q_i) \leq
v_{k-2}(q_j)$. Since $\weight'(q,(q,F)) \leq \weight'(q,(q,F'))$
if $F\subseteq F'$, we~have
\[
v_k(q) = \max_i \weight'(q,(q,\{q_i,\ldots,q_r\})) +
v_{k-2}(q_i). 
\]
Hence the sequence $v_{2k}$ can be computed in time $\Oh(k\cdot
\abs E)$ on~$\V{}$. Now, despite the exponential size of~$\G'$, the length
of a simple cycle in~$\G'$ is at most $2\abs{\V{}}$. Hence, Theorem~2.2
in~\cite{ZP96} becomes
\[
2k\cdot \val^{\G'}(q) - 4\abs{\V{}}\cdot W' \leq v_{2k}(q) \leq
  2k\cdot \val^{\G'}(q) +4\abs{\V{}}\cdot W'
\]
for all $q\in\V{}$,
where $W'$~is the maximal absolute weight in~$\G'$. Since
$W'\leq\abs{\V{}}\cdot 2W$, it follows from~\cite{ZP96} that
$\val^{\G}={-\val^{\G'}}\restrict\V{}$ can be computed in 
time $\Oh(\abs{\V{}}^4\cdot \abs E \cdot W)$.
\qed
\end{proof}
\fi

\iffull

Now, given a mean-penalty parity game~$\G$ with associated mean-payoff parity
game~$\G'$ and a set~$T$ of states of~$\G$, we~define
\begin{align*}
  \rebar[\G]{T} &= T\cup \{(q,F)\in \bar{\V{all}} \mid F\subseteq T\}; \\
  \drebar[\G]{T} &= T\cup \{(q,F)\in \bar{\V{all}} \mid F\cap T \not=\emptyset\}.
\end{align*}
We usually omit to mention the superscript~$\G$ when it is clear from the
context.

\begin{lemma}
If $S$~is a subarena of~$\G$, then $\rebar{S}$ and $\drebar{S}$ are
subarenas of~$\G'$. 
\end{lemma}

\begin{proof}
%
Assume that $S$~is a subarena of~$\G$, and pick a
state~$q$ in~$\rebar{S}$. If $q\in Q$, then it also
belongs to~$S$ and, as a state of~$\G$, has a
successor~$q'$ in~$S$. Then $\rebar{S}$ contains $(q,\{q'\})$, which
is a successor of~$q$. If $q$~belongs to~$\bar{\V{all}}$, then
$qE'\subseteq S$ by definition of $\rebar{S}$; hence it~has at least
one successor in~$S$. A~similar argument shows that $\drebar{S}$ is also a
subarena of~$\G'$. \qed
\end{proof}

\begin{lemma}\label{lemma-setbar}
Let $\G$ be a mean-penalty parity game
with associated mean-payoff parity 
game~$\G'$, and let $A,B\subseteq\V{all}$. Then
\begin{align*}
\rebar{A\cap B} &= \rebar A\cap \rebar B, &
  \rebar{A\cup B} &\supseteq \rebar A \cup \rebar B, \\
\drebar{A\cup B} &= \drebar A\cup \drebar B, &
  \drebar{A\cap B} &\subseteq \drebar A \cap \drebar B, \\
\rebar{\V{}\setminus A} &= \V{}'\setminus\drebar{A}, &
  \drebar{\V{}\setminus A} &= \V{}'\setminus\rebar{A}\,.
\end{align*}
\end{lemma}

\begin{proof}
Straightforward.\qed
\end{proof}

\begin{lemma}\label{lem:attr}\label{lemma-attr}
Let $\G$ be a mean-penalty parity game
with associated mean-payoff parity 
game~$\G'$, and let $F\subseteq\V{}$. Then
\begin{align*}
\rebar{\Attr_1^\G(F)} &=\Attr_1^{\G'}(F)=\Attr_1^{\G'}(\rebar{F}), \\
\drebar{\Attr_2^\G(F)} &=\Attr_2^{\G'}(F)=\Attr_2^{\G'}(\drebar{F})\,.
\end{align*}
\end{lemma}

\begin{proof}
We only prove the first statement; the second can be proved using similar
arguments. Clearly, $\Attr_1^{\G'}(F)=\Attr_1^{\G'}(\rebar{F})$,
so we only need to prove that $\rebar{\Attr^{\G}_1(F)}=\Attr_1^{\G'}(F)$.
First pick $q\in\rebar{\Attr^{\G}_1(F)}$. If $q\in\V{}$, then
the attractor strategy for reaching~$F$ can be mimicked in~$\G'$,
and therefore $q\in\Attr_1^{\G'}(F)$. On the other hand, if
$q\in\bar{\V{}}$, then all successors of~$q$ lie in $\Attr^{\G}_1(F)$
and therefore also in $\Attr_1^{\G'}(F)$. Hence, $q\in\Attr_1^{\G'}(F)$.
Now pick $q\in\Attr_1^{\G'}(F)$. If $q\in\V{}$, then the attractor strategy
for reaching~$F$ yields a multi-strategy~$\sigma$ in~$\G$ such that all
plays $\rho\in\Out^{\G}(\sigma,q)$ visit~$F$. Hence,
$q\in\Attr_1^{\G}(F)\subseteq\rebar{\Attr_1^{\G}(F)}$. On the other hand,
if $q\in\bar{\V{}}$, then all successors of~$q$ lie in
$\V{}\cap\Attr_1^{\G'}(F)$ (since $q$~is a \Pl2 state) and therefore also
in $\Attr_1^{\G}(F)$. Hence, $q\in\rebar{\Attr_1^{\G}(F)}$.\qed
\end{proof}

%

\fi

Algorithm~$\SSolve$ is our algorithm for computing the values
\begin{algorithm}
\vspace*{.8ex}
\begin{tabbing}
\hspace*{1em}\=\hspace{1em}\=\hspace{1em}\=\hspace{1em}\=
\hspace{1em}\=\hspace{1em}\= \kill

\textbf{Algorithm} $\SSolve(\G)$ \\[\medskipamount]

\emph{Input:} mean-penalty parity game $\G=(G,\col)$ \\
\emph{Output:} $\val^\G$ \\[\medskipamount]

\textbf{if} $\V{}=\emptyset$ \textbf{then return} $\emptyset$ \\
$p\coloneq\min\{\col(q)\mid q\in\V{}\}$ \\
\+\textbf{if} $p$~is even \textbf{then} \\
  $g\coloneq\SSolveMP(G,0)$ \\
  \textbf{if} $\col(q)=p$ for all $q\in\V{}$ \textbf{then return} $g$ \\
  $T\coloneq\V{}\setminus\Attr_1^{\G}(\col^{-1}(p))$;
  $f\coloneq\SSolve(\G\restrict T)$ \\
  $x\coloneq\max(f(T)\cup g(\V{}))$;
  $A\coloneq\Attr_2^{\G}(f^{-1}(x)\cup g^{-1}(x))$ \\
  \textbf{return} $(\V{}\to\bbR\cup\{\infty\}\colon q\mapsto x)\sqcap
    \SSolve(\G\restrict\V{}\setminus A)$ \-\\
\+\textbf{else} \\
  $T\coloneq\V{}\setminus\Attr_2^{\G}(\col^{-1}(p))$ \\
  \textbf{if} $T=\emptyset$ \textbf{then return}
    $(\V{}\to\bbR\cup\{\infty\}\colon q\mapsto\infty)$ \\
  $f\coloneq\SSolve(\G\restrict T)$; $x\coloneq\min f(T)$;
  $A\coloneq\Attr_1^{\G}(f^{-1}(x))$ \\
  \textbf{return} $(\V{}\to\bbR\cup\{\infty\}\colon q\mapsto x)\sqcup
    \SSolve(\G\restrict\V{}\setminus A)$ \-\\
\textbf{end if}
\end{tabbing}
\vspace*{-2ex}
\end{algorithm}
of a mean-penalty parity game. The algorithm employs as a subroutine
an algorithm $\SSolveMP$ for computing the values of a mean-penalty parity
with a trivial priority function (see Lemma~\ref{lemma-ZP}).
Since $\SSolveMP$ can be implemented to run in time
$\Oh(\abs{\V{}}^4\cdot\abs{E}\cdot W)$, the running time of
the procedure $\SSolve$ is $\Oh(\abs{\V{}}^{d+3}\cdot\abs{E}\cdot W)$.
Notably, the algorithm runs in polynomial time if the number of priorities
is bounded and we are only interested in the average \emph{number} of
edges blocked by a strategy in each step (\ie if all weights are
equal to~$1$).

\begin{theorem}\label{thm:mpep-det}
The values of a mean-penalty parity
game with $d$~priorities can be computed in time 
$\Oh(\abs{\V{}}^{d+3}\cdot \abs{E}\cdot W)$.
\end{theorem}

\iffull

\begin{proof}
From \cref{lemma-ZP} and with the same runtime analysis as in the
proof of \cref{thm:mpp-det}, we get that $\SSolve$ runs in time
$\Oh(\abs{\V{}}^{d+3}\cdot \abs{E}\cdot W)$. We now prove that the algorithm
is correct, by proving that there is a correspondence between the values
the algorithm computes on a mean-penalty parity game~$\G$ and the values
computed by Algorithm~$\Solve$ on the mean-payoff parity game~$\G'$.
More precisely, we show that $\Solve(\G')\restrict\V{}=-\SSolve(\G)$.
The correctness of the algorithm thus follows from
\cref{lem:value_equivalence}, which states that
${\val^{\G'}}\restrict\V{}=-\val^{\G}$.

The proof is by induction on the number of states in~$\G$.
The~result holds trivially if
$\V{}=\emptyset$. Otherwise, assume that the result is true for
all games with less than $\abs{\V{}}$~states and let
$p=\min\{\col(q)\mid q\in\V{}\}$. By
construction, $p$~is also the minimal priority in~$\G'$. We
only consider the case that $p$~is even; the other case is proved
using the same arguments.

Write $g'$, $T'$, $f'$, $x'$ and~$A'$ for the items
computed by $\SSolve$ on~$\G'$,
while $q$, $T$, $f$, $x$ and~$A$ are the corresponding items
computed by $\Solve$ on~$\G$. 
Then $g'(q)=-g(q)$ for all~$q\in\V{}$, and $g'((q,F))=\min_{q'\in F} g'(q')$
for all $(q,F)\in\bar{Q}$ (since such states belongs to \Pl2).
If $\G$~has only one priority, the result follows. Otherwise, by
\cref{lemma-setbar,lemma-attr}, we have $T'=\drebar T$.
However, any state~$(q,F)\in T'$ that is not a state of the
game $(\calG\restrict T)'$ has no predecessor in~$\G'\restrict T'$:
if $q\in T'$ then $q\in T\cap\V{1}$ and $qE\setminus T\neq\emptyset$,
\ie $qE\cap\Attr_1(\col^{-1}(p))\neq\emptyset$; but then
$q\in\Attr_1(\col^{-1}(p))$ and thus $q\notin T$, a contradiction.
It~follows that $\Solve(\G'\restrict T')\restrict
T=\Solve((\G\restrict T)')\restrict T$. 
  
Now, since $T$~is a strict subset of~$\V{}$, the induction hypothesis
applies,
so that $f'(t)=-f(t)$ for all~$t\in T$. It~follows that $x'=-x$.
Let $S\coloneq\V{}\setminus A$ and $S'\coloneq\V{}'\setminus A'$. By
\cref{lemma-attr}, $A'=\drebar A$, and by \cref{lemma-setbar},
$S'=\rebar{S}$. Again, any state
$(q,F)\in S'$ that is not a state of the
game $(\G\restrict S)'$ has no predecessor in
$\G'\restrict S'$. Hence,
$\Solve(\G'\restrict S')\restrict S
=\Solve((\G\restrict S)')\restrict S$
Applying the induction
hypothesis to the game $G\restrict S$, we~get that
$\Solve((\G\restrict S)')\restrict S=-\SSolve(G\restrict S)$,
and the result follows for~$\G$.
\qed
\end{proof}

\fi

\section{Conclusion}

In this paper, we have studied mean-payoff parity games, with an application
to finding permissive strategies in parity games with penalties. In particular,
we have established that mean-penalty parity games are not harder to solve than
mean-payoff parity games: for both kinds of games, the value problem is in
${\NP\cap\coNP}$ and can be solved by an exponential algorithm that becomes
pseudo-polynomial when the number of priorities is bounded.

One complication with both kinds of games is that optimal
strategies for \Pl1 require infinite memory, which makes it
hard to synthesise these strategies. A suitable alternative to
optimal strategies are \emph{$\epsilon$-optimal} strategies
that achieve the value of the game by at most~$\epsilon$.
Since finite-memory $\epsilon$-optimal strategies are
guaranteed to exist \cite{BCHJ09},
a challenge for future work is to modify our algorithms so that
they compute not only the values of the game but also a finite-memory
$\epsilon$-optimal (multi-)\dbr strategy for \Pl1.


\subsubsection*{Acknowledgement}

We thank an anonymous reviewer for pointing out the polynomial reduction
from mean-penalty parity games to mean-payoff parity games, which has
simplified the proof that mean-penalty parity games are in \NP.

\bibliographystyle{plain}
\bibliography{permissive}

\end{document}